\documentclass[letterpaper]{article}
\usepackage[T1]{fontenc}

\usepackage{geometry}
\usepackage{setspace}

\usepackage[style=chem-acs, bibencoding=utf8,backend=biber,natbib=true,doi=false]{biblatex}

\DeclareBibliographyDriver{article}{%
 \usebibmacro{bibindex}%
 \usebibmacro{begentry}%
 \printnames{author}%
 \newunit\newblock
 \printfield{title}%
 \newunit\newblock
 \usebibmacro{journal+issuetitle}%
 \newunit\newblock
 \usebibmacro{note+pages}%
 \newunit\newblock
 \usebibmacro{finentry}%
}

\usepackage{graphicx}
\usepackage{float}
\newfloat{scheme}{htbp}{los}
\floatname{scheme}{Scheme}
\floatname{chart}{Chart}
\newfloat{graph}{htbp}{loh}

\usepackage{chemformula} 
\usepackage[version = 4]{mhchem} 

\usepackage{amsmath}
\usepackage{amssymb}
\usepackage{dcolumn}
\usepackage{bm}
\usepackage{xcolor}

\usepackage{authblk}
\title{}
\author[1]{Marcel Strohmeier}
    \affil[1]{Department of Physics, University of Konstanz, 78457 Konstanz, Germany}
\author[2]{Samanwita Biswas}
    \affil[2]{Institute of Physics and Astronomy, University of Potsdam, 14476 Potsdam-Golm, Germany}
\author[1]{Wolfgang Belzig}
   \author[2]{Regina Hoffmann-Vogel}
  \author[1]{Elke Scheer}

\title{Spin-polarized transport in copper-oxide atomic junctions revealed by anomalous shot-noise behavior in presence of the Kondo effect}
\date{*Email: marcel.strohmeier@uni-konstanz.de}

\begin{document}
\doublespacing

\maketitle

\begin{abstract}
 \noindent 
 Noise measurements provide a valuable tool for revealing spin polarization effects in the electronic transport through quantum coherent conductors.  We present an extension of the Landauer description of shot noise to include energy-dependent transmission functions and apply it to explore local magnetic correlations in air-oxidized copper contacts, for which first-principle studies have predicted the emergence of ferromagnetic ground states, attributing certain atomic configurations with spin filtering capabilities. 
   By means of low-temperature transport measurements, we provide comprehensive experimental evidence, including hysteretic magnetoresistance and zero-bias anomalies (ZBAs) attributed to the Kondo effect, for the presence of local magnetism. The analysis of the anomalous shot noise in the presence of ZBAs allows us to determine the spin polarization of the current which may reach even full polarization, confirming the spin-filtering capability of copper oxide atomic contacts. 
 

\end{abstract}






\noindent 
Atomic-scale junctions formed from nonmagnetic metals provide a particularly appealing platform to explore how atomic-scale confinement and oxidation can profoundly modify magnetic and transport properties. Copper is a key material for electrical interconnects in the semiconductor industry, making a detailed understanding of its conductance properties essential. 
In the bulk, the monovalent Cu reveals nearly free-electron-like transport and is slightly diamagnetic, whereas Ni, the left neighbor in the periodic table, exhibits robust itinerant ferromagnetism. This contrast highlights the delicate balance between electronic structure and magnetism in transition metals. Notably, Cu can acquire magnetic character through oxidation, forming Cu(I) oxide and Cu(II) oxide. CuO$_x$ 
displays a rich variety of magnetic phases, ranging from bulk diamagnetism to complex multiferroic behavior \autocite{Bogenrieder2024, Kimura2008, Brown1991, Ain1992}. 
Ferromagnetic signatures have been observed in several nanoscale systems \autocite{Shih2008, Batsaikhan2020, Rehman2011,Gao2010, Quin2010, Gao2010_2, Das2022, Prabhakaran2013,Gao2010, Aiswarya2019, Quin2010, Gao2010_2, Chen2009, Soon2009, Prabhakaran2013,Batsaikhan2020, Shih2008, Das2022}. In many cases, non-stoichiometric surface magnetism has been proposed as a plausible explanation \autocite{Gao2010, Aiswarya2019, Quin2010, Gao2010_2, Chen2009, Soon2009, Prabhakaran2013}. Research on atomic-scale entities such as clusters or monoatomic chains, however, remains a largely unexplored field, where fundamental questions about the emergence of magnetism and its coupling to electronic transport persist.\\
Oxygen incorporation introduces partially filled $p$ orbitals and enhances hybridization with Cu $d$ states, creating favorable conditions for the formation of localized magnetic moments \autocite{Cakir2011}. First-principles calculations predict that Cu-O monoatomic chains can stabilize ferromagnetic ground states, exhibit strong 
SP, and act as efficient spin filters \autocite{Zheng2015}, despite the absence of magnetism in pristine Cu chains \autocite{Cakir2011}. Experimentally, oxygen assisted formation of elongated atomic chains in mechanically controllable break junctions has been observed, accompanied by a pronounced suppression of conductance relative to clean Cu contacts \autocite{Thijssen2008}. However, direct experimental evidence linking these structural motifs to magnetic order or correlated electron phenomena remains scarce, and the microscopic mechanisms governing spin dependent transport in Cu$_x$O atomic junctions are still poorly understood \autocite{Zheng2015, Thijssen2008, Chakrabarti2022}.

Here, we show that by combining the results of length dependent studies, magnetotransport studies transport spectroscopy and shot noise analysis allows to determine the spin polarization of ultimately small contacts. To this end, we derive the expressions for the conductance and the shot noise in the case of spin-polarized multichannel transport with energy-dependent transmission functions in the framework of the Landauer approach of quantum coherent transport. The comparison of the theoretical and experimental findings shows that zero-bias anomalies due to the Kondo effect and fully spin-polarized transport is possible in the same junctions.


For the fabrication and adjusting atomic-size contacts we apply the mechanically controllable break junction (MCBJ) method (see Fig.~\ref{fig:Histogram} (a) and the Supporting Information (SI)). The junctions are studied at low temperature (4.2 K) and under cryogenic vacuum conditions. First we study as-prepared Cu MCBJs to confirm the behavior known from clean metallic Cu atomic contacts \autocite{Thijssen2008, Strohmeier2023, Ludoph2000, Vardimon2016}. 
To investigate whether oxygen incorporation induces magnetic signatures in the electronic transport properties of atomic-scale Cu junctions, we then study MCBJs that were intentionally oxidized prior to low-temperature measurements (see SI).  

Already after a single rupture in air, pronounced changes in the contact formation are observed. Upon re-closing, the resistance often does not fully recover its initial metallic value and larger mechanical displacements
are required to re-establish contact. These observations indicate a modified atomic structure at the junction, consistent with partial oxidation of the Cu constriction due to surface reactions (see Fig.~\ref{fig:Histogram} (a)). In practice, this breaking and closing protocol can be repeated only a limited number of times (mostly two or three cycles), as excessive opening may prevent reliable contact reformation.


A first indication of oxygen incorporation is provided by conductance traces and histograms compiled from these. In the histograms shown in Fig.~\ref{fig:Histogram} (b), compared to pristine Cu junctions, air oxidized contacts exhibit a strong suppression of the characteristic conductance peaks near integer multiples of $G_0$, accompanied by an enhanced weight at low conductance values. This behavior reflects the stabilization of highly resistive atomic configurations and is consistent with earlier reports on oxygen assisted chain formation in Cu break junctions \autocite{Thijssen2008}.

Further insight is obtained from the analysis of individual opening traces and their associated chain lengths, which are depicted in Fig.~\ref{fig:Histogram} (d) and \ref{fig:Histogram} (c), respectively. After calibrating the electrode displacement (see Fig. S1 in the SI), the final stage of each rupture trace was analyzed by defining the chain length as the displacement between conductance values of 1.1\,$G_0$ and 0.05\,$G_0$. In oxidized junctions, the conductance evolution within this window is highly non-monotonic and varies strongly from trace to trace, indicative of a complex atomic rearrangement process involving Cu-O units. The chain length enhancement after oxidation is fully compatible with the report of Thijssen \emph{et al.} \autocite{Thijssen2008}.


Magnetotransport measurements were performed on mechanically fixed atomic Cu/Cu$_x$O junction configurations at $T \approx 4.2$\,K by sweeping a magnetic field perpendicular to the sample plane. The conductance $G = I/V$ was recorded continuously while ramping the field between $\pm B_\text{max}$, and only complete sweeps between positive and negative maximum fields were evaluated. Throughout all measurements, a constant bias voltage of 10\,mV was applied.

Fig.~\ref{fig:MC_records} summarizes the key experimental observations for air oxidized junctions. Panel (a) shows a collection of magnetoconductance (MC) traces exhibiting pronounced non-monotonic behavior as a function of magnetic field. Both positive and negative MC responses are observed, with relative conductance changes reaching several tens of percent. The curves are mainly symmetric upon field reversal but often hysteretic, and their shapes vary strongly from contact to contact. 
To facilitate comparison, the traces are vertically shifted and re-scaled.

For each MC trace, the corresponding current-voltage characteristics measured immediately prior to the field sweep are shown in panel (b). All d$I$/d$V$ - $V$ curves are smooth within the explored bias range, showing small variations on the scale of a few mV as typical for conductance fluctuations \autocite{Ludoph2000}, without pronounced zero-bias anomalies (ZBAs) or other spectroscopic signatures. 

The magnitude of the field induced conductance variation is quantified by the maximum magnetoconductance ratio
\begin{equation}
    \text{MCR}_\text{max} = \max\left(\frac{|G(B)-G(0)|}{G(0)}\right),
\end{equation}
extracted from each sweep. Figure~\ref{fig:MC_records} (c) displays $\text{MCR}_\text{max}$ as a function of the zero-field conductance for all continuous MC sweeps. Significant MC signals are predominantly found for junctions with conductances below $1\,G_0$, i.e.\ in a range corresponding to oxidized atomic configurations or short Cu-O chains. Both sweep directions yield comparable results, indicating that the extracted MCR values are robust. The MCR values and the field scale of the variation corresponds to the observations made for atomic contacts from the strong paramagnets Pt, Pd and Ir \autocite{Strigl2015, Strigl2016, Prestel2019} and indicate the emergence of local magnetic moments.

To exclude thermal artifacts due to eddy-current heating, Fig.~\ref{fig:MC_records} (d) shows a representative MC trace together with the simultaneously recorded sample temperature. The temperature remains constant during the entire magnetic-field sweep, demonstrating that the observed conductance variations cannot be attributed to field induced heating or temperature drifts. Combined with the absence of abrupt conductance jumps, 
magnetostriction or thermal effects are ruled out as the dominant origin of the MC features.


Signatures of magnetic order may also occur as spectroscopic features in current-voltage characteristics. The most prominent one stems from the Kondo effect, a many-body phenomenon which was first discovered in metals containing dilute magnetic impurities and manifests itself in nanocontacts as a ZBA in the differential conductance. At the atomic scale, such resonances commonly acquire a Fano line shape due to quantum interference between a localized magnetic moment and a continuum of conduction channels \autocite{Calvo2009}. A sketch of such a scenario, with exemplary Fano line-shaped transmission functions, is illustrated in Fig.~\ref{fig:Kondo_MS_EM2_09}(a).

In oxidized Cu break junctions, weak ZBAs with 
conductance values well below $1 \,G_0$
are observed in a subset of atomic contacts.
Representative examples of d$I$/d$V$ spectra are shown in Fig.~\ref{fig:Kondo_MS_EM2_09} (b,c), together with fits to the standard Fano expression \autocite{Calvo2012}
\begin{equation}
\frac{\text{d}I}{\text{d}V}(V)
= g_0
+ A \,
\frac{(q + \varepsilon)^2}{1 + \varepsilon^2},
\qquad
\varepsilon = \frac{eV - \varepsilon_K}{\Gamma_K}\;.
\label{eq:fano_line_shape}
\end{equation}

From these fits, the resonance amplitude $A$, shape factor $q$, and the Kondo temperature $T_K$ are extracted. With $\Gamma_K = k_B T_K$ the Kondo parameters can be further linked to the impurity level occupation \autocite{Calvo2012}

\begin{equation}
    n_d = 1 - \frac{2}{\pi} \arctan \left( \frac{\varepsilon_K}{k_B T_K} \right)\;.
\end{equation}

A statistical analysis of all spectra with $A>0.01\,G_0$ reveals a broad distribution of Kondo temperatures (Fig.~\ref{fig:Kondo_MS_EM2_09} (d)), which is well described by a log-normal form, reflecting the exponential dependence of $T_K$ on the microscopic parameters of the junction. The mean value, $\overline{T}_K \approx 150$\,K, lies in the range reported for atomic-scale Kondo systems in transition metal contacts. The inferred impurity occupation is centered close to unity, consistent with a predominantly singly occupied magnetic level.

In Fig.~\ref{fig:Kondo_MS_EM2_09} (e) we plot $A$ vs. $T_K$  revealing no systematic correlation between both parameters. The 
amplitudes $A$ are similar to those reported for ferromagnetic nanocontacts \autocite{Calvo2012}, hence suggesting local magnetic order. 

The distribution of Fano parameters shows a strong preference for dip-like or weakly asymmetric resonances with $|q|\lesssim 1$ [Fig.~\ref{fig:Kondo_MS_EM2_09} (f)]. This indicates that transport through Cu$_x$O atomic contacts is dominated by a background channel i.e., with $s$-level character, while the Kondo resonance contributes mainly via destructive interference.

Overall, the observed Fano line shapes provide spectroscopic evidence for localized magnetic moments in oxidized Cu atomic junctions for a small subset (1-2 \%) of the junctions. While most ZBAs are weak, they establish a direct link between oxygen incorporation and many-body spin correlations, forming the basis for the anomalous shot noise behavior discussed in the following section. 
We note that in metallic Cu contacts (studied in cool-downs before oxidation) no such ZBAs were observed. However, it has to be taken into account that the probability of contact formation with conductance values below $1 G_0$ is very low for Cu contacts.


 In addition to MC 
measurements, we study the shot noise of atomic contacts at $B = 0$. These non-equilibrium current fluctuations the amplitude of which is proportional to the current for quantum coherent conductors with linear current-voltage characteristics provide direct access to the number of conduction channels, their transmission probabilities, and possible SP  \autocite{vandenBrom1999, Vardimon2013, Burtzlaff2015}. The noise measurement setup is shown in Fig.~\ref{fig:noise_spinpol_contacts} (b) and follows the same design as in our previous study \autocite{Prestel2021}. It is optimized for direct shot noise detection using a current amplifier with an extended bandwidth up to 800\,kHz.


Being in the quantum-coherent regime, the shot noise is first analyzed within the Landauer formalism. For a system at finite temperature $T$, the corresponding expression for the noise power reads \autocite{Blanter2000}

\begin{align}\label{eq:full_shot_noise}
S_I &= e V G_0 \coth\!\left( \frac{eV}{2 k_B T} \right) 
\sum_{i, \sigma} \tau_{i, \sigma} (1 - \tau_{i, \sigma}) \\
&+ 2 k_B T G_0 \sum_{i, \sigma} \tau_{i, \sigma}^2 , \notag
\end{align}
where $V$ is the applied bias voltage and $k_B$ denotes the Boltzmann constant. To assess the Fano factor $F = {S_I}/{2 e \langle I \rangle} = {\sum_{i, \sigma} \tau_{i, \sigma} (1 - \tau_{i, \sigma})}/{\sum_{i, \sigma} \tau_{i, \sigma}}$ and, hence, the degree of SP
\begin{equation}
    SP = \dfrac{\left| G_\uparrow - G_\downarrow \right|}{G_\uparrow + G_\downarrow}\;,
\end{equation}
we follow the methodology of Kumar \emph{et al.} \autocite{Kumar2012} by introducing the normalized noise $Y = \left[ S_I(V) - S_I(0) \right] /S_I(0)$ and normalized voltage $X = x\,\coth(x)$ with $x = eV / 2 k_B T$. Handling the data via the linear relation
\begin{equation}\label{eq:1_f_fit}
    Y = F \left( X - 1 \right)
\end{equation} 
allows $F$ to be obtained directly, without explicitly determining the transmission configuration $\{ \tau_{i,\sigma} \}$.

To obtain statistically meaningful results, ensembles of atomic contacts were analyzed across four different samples. For the purpose of probing spin filtering signatures in Cu$_x$O junctions [cf. Fig.~\ref{fig:noise_spinpol_contacts} (a)], we are primarily interested in the conductance range below $1\,G_0$ 
\autocite{Cakir2011, Zheng2015}, which is 
the regime where the Fano factor analysis allows the detection of a potential SP (see Fig.~\ref{fig:noise_spinpol_contacts} (c)) - Does it require a more detailed description?. Noise spectra were acquired for each configuration at discrete bias voltages 
and analyzed following the procedure that is outlined in 
\autocite{Prestel2021} (see Fig. S2 in the SI for detailed information).

For all junctions exhibiting a sufficiently constant d$I$/d$V$ characteristics 
over the bias range of the noise measurements, we focus on their representation in the $F(G)$ space. The analysis as depicted in Fig.~\ref{fig:noise_spinpol_contacts} (d) shows that the majority of data points lies above the spin degenerate single-channel limit (black solid line), indicating that charge transport involves more than one spin degenerate channel. In this situation, the Fano factor does not allow conclusions about spin polarized currents since any enhancement of $F$, whether due to a multichannel situation or orbital degeneracy, would overcompensate the noise reduction induced by SP \autocite{Prestel2021}. Our investigation shows, however, that a minute fraction of contacts falls indeed within the regime that is not compatible with transport through a single, spin degenerate channel. 
We present two examples in Fig.~\ref{fig:noise_spinpol_contacts}, where the noise spectra for each contact are shown in (e) and the corresponding noise-voltage $Y(X)$ are given in (f). In both cases, we observe a partial SP of about 60\% and 45\%, as illustrated by the two red star-filled data points in (d). Across all investigated samples, the extracted SP value varies between 40\% and 100\%, indicating that the degree of spin selectivity can differ substantially among different Cu$_x$O junctions. 


As a next step, we go beyond the conventional Fano factor analysis, the interpretation of which is more involved 
by the presence of at least two contributing channels. The focus is set on contacts with strongly nonlinear $I-V$ characteristics featuring a pronounced ZBA. 
Despite the fact that our measurement procedure does not allow for temperature or magnetic-field dependent studies, there are reasonable arguments to interpret spectroscopic features, as exemplary shown in Fig.~\ref{fig:noise_model} (a), as an emerging Kondo resonance. On contacts of this type, the excess noise $S_{I,ex} (V) = S_I(V)-S_I(0)$ exhibits a nonlinear dependence on the bias voltage, which cannot be simply captured by Eq.~\eqref{eq:full_shot_noise}. 

In the classical Landauer picture, transport properties are typically treated within the linear-response regime, implying that $G$ and $S_I$ depend on constant transmission coefficients. If this assumption fails, it becomes necessary to incorporate non-linearities into an extended model. To link both observations, the bias dependence of $G$ and the noise, we build on the work of Tewari~\emph{et al.} \autocite{Tewari2018} and Mu~\emph{et al.} \autocite{Mu2019}, who reported on the anomalous shot noise arising in atomic-scale junctions with nontrivial potential barriers 
Allowing the transmission function $T(E)$ to be energy dependent provides a natural way to account for nonlinear bias scaling in both the current and its non-equilibrium fluctuations. Their specific functional behavior depends on $T(E)=\sum_{i,\sigma}\tau_{i,\sigma}(E)$, which in general may consist of several energy-dependent spin channel contributions. In the zero-temperature limit, the shot noise can then be written as
\begin{align}
    S_{I}(V)
    &= G_0
    \sum_{i,\sigma}\;\int_{-eV/2}^{eV/2}
     \tau_{i,\sigma}(E)\,\big[1-\tau_{i,\sigma}(E)\big]\, \text{d}E 
\end{align}
where the integration extends over the energy window $eV$ set by the potential difference $V$. To determine the transmission function experimentally, we start from the measured $I-V$ characteristics and their numerical derivative. The differential conductance is fitted by the Fano formula given in Eq.~\eqref{eq:fano_line_shape}, yielding an analytical function $\left(\text{d}I/\text{d}V\right)_\text{fit}(V)$, as exemplary shown by the red line in Fig.~\ref{fig:noise_model} (a). After the successful parametrization and, hence, capturing the physics of the Kondo related ZBA, the resulting total transmission
\begin{equation}
    T(E) = \dfrac{1}{G_0}\left(\dfrac{\text{d}I}{\text{d}V}\right)_\text{fit}(V)
\end{equation}
serves as the basis for the subsequent modeling of the noise characteristics. As shown in Fig.~\ref{fig:noise_model} (b), interpreting $T(E)$ as a single spin degenerate channel leads to calculated excess noise that exceeds the measured data. 
Similarly, the conductance measurements suggest that more than one spin channel is active in order to reach conductances larger than 0.5\,$G_0$. 
To empirically account for all measured Cu$_x$O junctions showing anomalous shot noise in the presence of a strong zero-bias peak (see  Fig. S3 and S4 in the SI for more examples), we construct a minimal two-channel model (2CM), which is sketched in Fig.~\ref{fig:noise_model} (e). In this model, one channel $\tau_1(E)$ is spin polarized and carries the energy dependent profile of the Kondo resonance, while the second channel $\tau_0$ is spin degenerate and energy independent, providing a constant background contribution. As discussed 
below, the SP may arise from the apex atoms, interconnecting the macroscopic leads with the central constriction. Their full transmission is   
\begin{align}
    T(E) = 2\tau_{0}+\tau_{1,\uparrow}(E)+\tau_{1,\downarrow}(E)
\end{align}  
and the zero-temperature shot noise is obtained by summing the spin resolved contributions of each channel:
\begin{align}\label{Eq:2CM_noise}
    S_{I,\text{2CM}}(V) &= G_0 \sum_{\sigma=\uparrow,\downarrow} \int_{-eV/2}^{eV/2} \tau_{1,\sigma}(E)\,[1-\tau_{1,\sigma}(E)]\,\mathrm{d}E \notag \\
    &\quad + 2 e V G_0 \, \tau_0 (1-\tau_0).
\end{align}  

For a more detailed derivation of Eq.~\eqref{Eq:2CM_noise} and the technical procedure behind concrete calculations, the reader is referred to the SI. We parametrize the 2CM by two parameters: (1) the SP of the Fano active channel with
\begin{equation}\label{eq:2CM_SP}
    SP = \frac{|\tau_{1,\uparrow} - \tau_{1,\downarrow}|}{\tau_{1,\uparrow} + \tau_{1,\downarrow}},
\end{equation}  
and (2) the relative contribution of the two channels, captured by the transmission ratio
\begin{equation}\label{eq:2CM_R}
    R = \frac{\tau_1(E\gg k_B T_K)}{2\tau_0}.
\end{equation}  
By adjusting $SP$ and $R$, the model quantitatively reproduces the measured excess noise versus bias for all contacts, as explicitly shown by the red line in Fig.~\ref{fig:noise_model} (b). The corresponding transmission functions underlying the 2CM for $SP=100\%$ and $R=3.13$ are shown in Fig.~\ref{fig:noise_model} (c). Additional insight into the contact configuration is provided by the contact history in Fig.~\ref{fig:noise_model} (d), where the conductance, plotted as a function of electrode displacement, indicates the formation of a short atomic chain of approximately 2\,\AA\;during the noise measurements. Upon further elongation, the junction can be stretched by additional $\sim$2.5\,\AA\;before entering the tunneling regime. 

The total displacement, together with the non-monotonic conductance evolution during stretching, supports the conclusion that oxygen is involved in the contact formation. 

Finally, after having provided the above additional argument 
for the involvement of two conduction channels, we revisit the Fano factor analysis of all junctions and interpret it within the notation of the 2CM. For fixed combinations of $SP$ and $R$ and assuming $G=G_0 T(E)$, an array of $F(G)$ curves can be calculated. In the range $G < G_0$, these curves form straight lines, such as the one indicated by the colored marker symbols in Fig.~\ref{fig:noise_model} (f), which confirm the expected anticorrelation between $F$ and SP. Most data points lie below the upper limit of maximum noise (gray line), where the conductance would be carried by two equivalent spin degenerate channels
These observations consistently underline the two-channel picture in Cu$_x$O atomic junction.


 Our experimental findings suggest that the transport properties of oxidized Cu break junctions are governed by Cu-O entities forming the narrowest part of the constriction. While the macroscopic Cu electrodes can be treated as diamagnetic reservoirs, surface oxidation is expected to become more dominant closer to
the constriction, where the conductor narrows down to the nanoscale. In a minimum model as schematically illustrated in Fig.~\ref{fig:drawing_contact_model}, non-stoichiometric effects and surface magnetism may induce a finite magnetization in the vicinity of the apex atoms (light gray atoms). Such magnetic moments (gray arrows) are known to couple to the conduction electrons, with the capacity to act as spin polarizers or to generate non-monotonic MC signals.

Following theoretical predictions \autocite{Cakir2011}, we assume that Cu-O chains host ferromagnetic ground states, giving rise to a net magnetic moment (black arrow). For geometries with at least two incorporated oxygen atoms, this ground state also leads to spin polarized transmission channels. The dominant contributions arise from hybridized O(2$p$) and Cu(3$d$) orbitals, which form partially transmitting channels near the Fermi level \autocite{Zheng2015}.

Motivated by the Fano factor statistics and the Kondo analysis, a minimal transport description must include at least two effective channels. From an experimental perspective, the precise orbital composition of the transport channels -- i.e., the relative contributions of $s$, $p$ and $d$ states -- cannot be determined unambiguously. Nonetheless, all relevant orbitals should be regarded as potential contributors to transport. 

As the analysis of the noise measurements in the presence of Kondo anomalies clearly shows, the channel with the energy dependent transmission is in most cases spin-polarized, up to full SP. This fact is as first glimpse unexpected since experimental and theoretical studies on quantum dots connected with ferromagnetic leads clearly demonstrated \autocite{Pasupathy2004, Martinek2003}. In these studies a split Kondo resonance was observed. To reconcile our findings of a non-split ZBA with the picture of Kondo physics we argue that the local magnetic fields acting on the conduction electrons might be too small to result in a measurable broadening of the ZBA. We estimate the strength of the local magnetic field by the hysteresis of the magnetoconductance curves which is in the order of 150\,mT. The $T_\text{K}$  of the junctions revealing  unconventional noise features is in the order of at least 20\,K, when fitting with the conventional spin degenerate theory \autocite{Costi2000}. For such high $T_\text{K}$, a broadening would be expected to be observable for magnetic fields in the order of at least 3\,T. In fact, the assumption of spin-degeneracy might have resulted in overestimating $T_K$. However, also for $T_\text{K}$s in the range of our measurement temperature of $\approx 4.2\,K$, such small magnetic fields would have a negligible influence. Even lower $T_\text{K}$s would not have been detectable in our experiment.    


In conclusion, we investigated the transport properties of air oxidized Cu MCBJs using a combination of magnetotransport, d$I$/d$V$ spectroscopy, and shot noise measurements. The experiments consistently point toward the emergence of local magnetic moments in Cu$_x$O atomic contacts as predicted by theory \autocite{Cakir2011, Zheng2015}. Specifically, the non-monotonic MC signals with large relative amplitudes indicate the presence of magnetic scattering centers in and around the constriction. The detection of ZBA with Fano line shapes complements this picture as it underscores the possible formation of Kondo resonances associated with the localized Cu$_x$O moments.

Shot noise spectroscopy provides complementary information on the underlying transport channels. While the Fano factor analysis of most contacts below one $G_0$ highlights their multichannel nature, it also reveals, through a small fraction of contacts compatible with a single channel, that (partially) spin polarized currents may occur. For junctions with strong ZBA, the excess noise shows an anomalous bias dependence that goes beyond the linear Landauer formalism. We introduce a minimal two-channel model which acquires an energy dependent transmission function that carries the profile of the Kondo resonance and, hence, requires a nonlinear noise modeling. Comparing the model based calculation with our experimental data, similarly suggests the presence of spin polarization, which, remarkably, coexists with the Kondo correlations. We interpret this coexistence as another hint for the emergence of a local magnetization 
in the nanoscale vicinity of the apex atoms, likely induced by surface oxidation. 
Taken together, these experimental findings provide a first coherent demonstration that oxygen incorporation into Cu nanocontacts can indeed induce localized magnetic moments and spin dependent transport at the atomic scale.

\section*{Acknowledgments}
We thank Theodoulos Costi and Oren Tal for helpful discussion. We gratefully acknowledge the nano.lab of the University of Konstanz for access to its equipment and its staff, Matthias Hagner and Annika Zuschlag, for their expert advice. We gratefully acknowledge financial support from the Deutsche Forschungsgemeinschaft (DFG; German Research Foundation) via SFB 1432 (project No. 425217212) and under project No. 493158779
as part of the collaborative research project SFB F 86 Q-M\&S funded by the Austrian Science Fund (FWF, project number LAP 8610-N).

\section*{Associated Content}

The the supporting information file \verb|CuO_Supporting_information.pdf| 
is available free of charge.
  It contains information about sample fabrication and measurement setup,  the calibration of the displacement axis of the junction,  the noise measurement setup and modeling, more examples for junctions revealing a zero-bias anomaly and unconventional noise spectra as well as additional details about the derivation of the two-channel transport model.

\section*{Author Information}

Corresponding author: Marcel Strohmeier (marcel.strohmeier@uni-konstanz.de).\\ Author contributions: MS performed the experiments and the data analysis. All authors discussed the results. MS wrote the manuscript with input from all authors. The authors declare no conflict of interest.
\printbibliography

@article{Pasupathy2004,
    author       = {Pasupathy, A. N. and Bialczak, R. C. and Martinek, J. and Grose, J. E. and Donev, L. A. K. and McEuen, P. L. and Ralph, D. C.},
    year         = {2004},
    title        = {The Kondo effect in the presence of ferromagnetism},
    volume       = {306},
    pages        = {86},
    journal      = {Science},
    publisher    = {{American Association for the Advancement of Science (AAAS)}},
    issn         = {0036-8075},
    language     = {english},
  doi = {10.1126/science.1102068}
  }

@article{Martinek2003,
    author       = {Martinek, J. and Utsumi, Y. and Imamura, H. and Barnas, J. and Maekawa, S. and Schön, Gerd},
    year         = {2003},
    title        = {Kondo effect in quantum dots coupled to ferromagnetic electrodes},
    volume       = {18},
    number       = {1},
    pages        = {75-76},
    journal      = {Physica / E},
    publisher    = {{North-Holland Publishing}},
    issn         = {1386-9477},
    language     = {english},
doi = {10.1103/PhysRevLett.91.127203},
url = {https://doi.org/10.1103/PhysRevLett.91.127203}
}

@article{Costi2000,
    author       = {T. A. Costi},
    year         = {2003},
    title        = {Kondo Effect in a Magnetic Field and the Magnetoresistivity of Kondo Alloys},
    volume       = {85},
    pages        = {1504--1507},
    journal      = {Phys. Rev. Lett.},
    publisher    = {American Physical Society},
    language     = {english},
doi = {10.1103/PhysRevLett.85.1504},
url = {DOI: https://doi.org/10.1103/PhysRevLett.85.1504}
}

@article{Cakir2011,
  title = {Effect of impurities on the mechanical and electronic properties of {Au}, {Ag}, and {Cu} monatomic chain nanowires},
  author = {\ifmmode \mbox{\c{C}}\else \c{C}\fi{}ak\ifmmode \imath \else\i \fi{}r, D. and G\"ulseren, O.},
  journal = {Phys. Rev. B},
  volume = {84},
  issue = {8},
  pages = {085450},
  numpages = {10},
  year = {2011},
  month = {08},
  publisher = {American Physical Society},
  doi = {10.1103/PhysRevB.84.085450},
  url = {https://link.aps.org/doi/10.1103/PhysRevB.84.085450}
}

@article{Tewari2018,
author = {Tewari, Sumit and van Ruitenbeek, Jan},
title = {Anomalous Nonlinear Shot Noise at High Voltage Bias},
journal = {Nano Letters},
volume = {18},
number = {8},
pages = {5217-5223},
year = {2018},
doi = {10.1021/acs.nanolett.8b02176},
URL = {https://doi.org/10.1021/acs.nanolett.8b02176},
}

@article{Mu2019,
author = {Mu, Anqi and Shein-Lumbroso, Ofir and Tal, Oren and Segal, Dvira},
title = {Origin of the Anomalous Electronic Shot Noise in Atomic-Scale Junctions},
journal = {The Journal of Physical Chemistry C},
volume = {123},
number = {39},
pages = {23853-23862},
year = {2019},
doi = {10.1021/acs.jpcc.9b06766},
URL = {https://doi.org/10.1021/acs.jpcc.9b06766},
}

@article{Ludoph2000,
  title = {Conductance fluctuations as a tool for investigating the quantum modes in atomic-size metallic contacts},
  author = {Ludoph, B. and Ruitenbeek, J. M van},
  journal = {Phys. Rev. B},
  volume = {61},
  issue = {3},
  pages = {2273--2285},
  numpages = {0},
  year = {2000},
  month = {01},
  publisher = {American Physical Society},
  doi = {10.1103/PhysRevB.61.2273},
  url = {https://link.aps.org/doi/10.1103/PhysRevB.61.2273}
}

@article{Zheng2015,
    author = {Zheng, Xiaolong and Xie, Yi-Qun and Ye, Xiang and Ke, San-Huang},
    title = {Conductance and spin-filter effects of oxygen-incorporated {Au}, {Cu}, and {Fe} single-atom chains},
    journal = {Journal of Applied Physics},
    volume = {117},
    number = {4},
    pages = {043902},
    year = {2015},
    month = {01},
    issn = {0021-8979},
    doi = {10.1063/1.4906439},
    url = {https://doi.org/10.1063/1.4906439},
}

@article{Strigl2015,
    title = {Emerging magnetic order in platinum atomic contacts and chains},
    author = {Strigl, Florian and Espy, Christopher and B\"uckle, Maximilian and Scheer, Elke and Pietsch, Torsten},
    journal = {Nature Comm.},
    volume = {6},
    pages = {6172},
    year = {2015},
    doi = {https://doi.org/10.1038/ncomms7172},
    url = {https://doi.org/10.1038/ncomms7172},
}

@article{Vardimon2013,
  title = {Experimental determination of conduction channels in atomic-scale conductors based on shot noise measurements},
  author = {Vardimon, Ran and Klionsky, Marina and Tal, Oren},
  journal = {Phys. Rev. B},
  volume = {88},
  issue = {16},
  pages = {161404},
  numpages = {5},
  year = {2013},
  month = {10},
  publisher = {American Physical Society},
  doi = {10.1103/PhysRevB.88.161404},
  url = {https://link.aps.org/doi/10.1103/PhysRevB.88.161404}
}

@article{Burtzlaff2015,
  title = {Shot Noise as a Probe of Spin-Polarized Transport through Single Atoms},
  author = {Burtzlaff, Andreas and Weismann, Alexander and Brandbyge, Mads and Berndt, Richard},
  journal = {Phys. Rev. Lett.},
  volume = {114},
  issue = {1},
  pages = {016602},
  numpages = {5},
  year = {2015},
  month = {01},
  publisher = {American Physical Society},
  doi = {10.1103/PhysRevLett.114.016602},
  url = {https://link.aps.org/doi/10.1103/PhysRevLett.114.016602}
}

@article{vandenBrom1999,
  title = {Quantum Suppression of Shot Noise in Atom-Size Metallic Contacts},
  author = {van den Brom, H. E. and van Ruitenbeek, J. M.},
  journal = {Phys. Rev. Lett.},
  volume = {82},
  issue = {7},
  pages = {1526--1529},
  numpages = {0},
  year = {1999},
  month = {02},
  publisher = {American Physical Society},
  doi = {10.1103/PhysRevLett.82.1526},
  url = {https://link.aps.org/doi/10.1103/PhysRevLett.82.1526}
}

@article{Vardimon2016,
  title = {Orbital origin of the electrical conduction in ferromagnetic atomic-size contacts: Insights from shot noise measurements and theoretical simulations},
  author = {Vardimon, R. and Matt, M. and Nielaba, P. and Cuevas, J. C. and Tal, O.},
  journal = {Phys. Rev. B},
  volume = {93},
  issue = {8},
  pages = {085439},
  numpages = {15},
  year = {2016},
  month = {02},
  publisher = {American Physical Society},
  doi = {10.1103/PhysRevB.93.085439},
  url = {https://link.aps.org/doi/10.1103/PhysRevB.93.085439}
}

@article{Ruitenbeek1996,
    author = {van Ruitenbeek, J. M. and Alvarez, A. and Piñeyro, I. and Grahmann, C. and Joyez, P. and Devoret, M. H. and Esteve, D. and Urbina, C.},
    title = {Adjustable nanofabricated atomic size contacts},
    journal = {Review of Scientific Instruments},
    volume = {67},
    number = {1},
    pages = {108-111},
    year = {1996},
    month = {01},
    issn = {0034-6748},
    doi = {10.1063/1.1146558},
    url = {https://doi.org/10.1063/1.1146558},
}

@article{Prestel2021,
  title = {Revealing channel polarization of atomic contacts of ferromagnets and strong paramagnets by shot-noise measurements},
  author = {Prestel, Martin W. and Strohmeier, Marcel and Belzig, Wolfgang and Scheer, Elke},
  journal = {Phys. Rev. B},
  volume = {104},
  issue = {11},
  pages = {115434},
  numpages = {10},
  year = {2021},
  month = {09},
  publisher = {American Physical Society},
  doi = {10.1103/PhysRevB.104.115434},
  url = {https://link.aps.org/doi/10.1103/PhysRevB.104.115434}
}

@article{Chakrabarti2022,
  title={Magnetic control over the fundamental structure of atomic wires},
  author={Chakrabarti, Sudipto and Vilan, Ayelet and Deutch, Gai and Oz, Annabelle and Hod, Oded and Peralta, Juan E and Tal, Oren},
  journal={Nature Comm.},
  volume={13},
  number={1},
  pages={4113},
  year={2022},
  publisher={Nature Publishing Group UK London},
  url = {https://doi.org/10.1038/s41467-022-31456-4}
}

@article{Thijssen2008,
doi = {10.1088/1367-2630/10/3/033005},
url = {https://dx.doi.org/10.1088/1367-2630/10/3/033005},
year = {2008},
month = {03},
publisher = {},
volume = {10},
number = {3},
pages = {033005},
author = {Thijssen, W H A and Strange, M and aan de Brugh, J M J and van Ruitenbeek, J M},
title = {Formation and properties of metal–oxygen atomic chains},
journal = {New Journal of Physics}
}

@article{Ain1992,
doi = {10.1088/0953-8984/4/23/009},
url = {https://dx.doi.org/10.1088/0953-8984/4/23/009},
year = {1992},
month = {06},
publisher = {},
volume = {4},
number = {23},
pages = {5327},
author = {M Ain and A Menelle and B M Wanklyn and E F Bertaut},
title = {Magnetic structure of {CuO} by neutron diffraction with polarization analysis},
journal = {Journal of Physics: Condensed Matter},
}

@article{Brown1991,
doi = {10.1088/0953-8984/3/23/016},
url = {https://dx.doi.org/10.1088/0953-8984/3/23/016},
year = {1991},
month = {06},
publisher = {},
volume = {3},
number = {23},
pages = {4281},
author = {P J Brown and T Chattopadhyay and J B Forsyth and V Nunez},
title = {Antiferromagnetism in {CuO} studied by neutron polarimetry},
journal = {Journal of Physics: Condensed Matter},
}

@article{Kimura2008,
  title={Cupric oxide as an induced-multiferroic with high-{T}$_C$},
  author={Kimura, T and Sekio, Y and Nakamura, H and Siegrist, T and Ramirez, AP},
  journal={Nature Materials},
  volume={7},
  number={4},
  pages={291--294},
  year={2008},
  publisher={Nature Publishing Group UK London},
  url = {https://doi.org/10.1038/nmat2125}
}

@article{Rehman2011,
  title={Size effects on the magnetic and optical properties of {CuO} nanoparticles},
  author={Rehman, Shama and Mumtaz, A and Hasanain, SK},
  journal={Journal of Nanoparticle Research},
  volume={13},
  pages={2497--2507},
  year={2011},
  publisher={Springer},
  url = {https://doi.org/10.1007/s11051-010-0143-8} 
}

@article{Chen2009,
doi = {10.1088/0953-8984/21/14/145601},
url = {https://dx.doi.org/10.1088/0953-8984/21/14/145601},
year = {2009},
month = {03},
publisher = {},
volume = {21},
number = {14},
pages = {145601},
author = {Chen, Chinping and He, Lin and Lai, Lin and Zhang, Hua and Lu, Jing and Guo, Lin and Li, Yadong},
title = {Magnetic properties of undoped
{Cu}$_2${O} 
fine powders with magnetic impurities and/or cation vacancies},
journal = {Journal of Physics: Condensed Matter},
}

@article{Shih2008,
    author = {Shih, Po-Hsun and Ji, Jhong-Yi and Ma, Yuan-Ron and Wu, Sheng Yun},
    title = {Size effect of surface magnetic anisotropy in {Cu}$_2${O} nanoparticles},
    journal = {Journal of Applied Physics},
    volume = {103},
    number = {7},
    pages = {07B735},
    year = {2008},
    month = {03},
    issn = {0021-8979},
    doi = {10.1063/1.2839331},
    url = {https://doi.org/10.1063/1.2839331},
}

@article{Prabhakaran2013,
title = {Room temperature ferromagnetic properties of {Cu}$_2${O} microcrystals},
journal = {Journal of Alloys and Compounds},
volume = {579},
pages = {572-575},
year = {2013},
issn = {0925-8388},
doi = {https://doi.org/10.1016/j.jallcom.2013.07.094},
url = {https://www.sciencedirect.com/science/article/pii/S0925838813017192},
author = {G. Prabhakaran and Ramaswamy Murugan},
}

@article{Das2022,
title = {Suppression of ferromagnetic order in {CuO}/{Cu}$_2${O} nanocomposites},
journal = {Materials Today Communications},
volume = {32},
pages = {104038},
year = {2022},
issn = {2352-4928},
doi = {https://doi.org/10.1016/j.mtcomm.2022.104038},
url = {https://www.sciencedirect.com/science/article/pii/S2352492822008911},
author = {R. Das and J. Alonso and E.M. Jefremovas and L. {Fernández Barquín} and P.K. Ngoc and H.T. Nguyen and D.T. Viet and P.V. Vinh and A.T. Duong},
}

@article{Gao2010_2,
  title={Vacancy-mediated magnetism in pure copper oxide nanoparticles},
  author={Gao, Daqiang and Zhang, Jing and Zhu, Jingyi and Qi, Jing and Zhang, Zhaohui and Sui, Wenbo and Shi, Huigang and Xue, Desheng},
  journal={Nanoscale research letters},
  volume={5},
  pages={769--772},
  year={2010},
  publisher={Springer},
  url = {https://doi.org/10.1007/s11671-010-9555-8}
}

@article{Quin2010,
title = {Room-temperature ferromagnetism in {CuO} sol–gel powders and films},
journal = {Journal of Magnetism and Magnetic Materials},
volume = {322},
number = {14},
pages = {1994-1998},
year = {2010},
issn = {0304-8853},
doi = {https://doi.org/10.1016/j.jmmm.2010.01.021},
url = {https://www.sciencedirect.com/science/article/pii/S0304885310000399},
author = {Hongwei Qin and Zhongli Zhang and Xing Liu and Yongjia Zhang and Jifan Hu},
}

@article{Aiswarya2019,
title = {Nanostructured {CuO} with antiferromagnetic core and weakly ferromagnetic shell},
journal = {Journal of Solid State Chemistry},
volume = {278},
pages = {120911},
year = {2019},
issn = {0022-4596},
doi = {https://doi.org/10.1016/j.jssc.2019.120911},
url = {https://www.sciencedirect.com/science/article/pii/S0022459619304165},
author = {A.S. {Aiswarya Raj} and G. Madhu and V. Biju},
}

@article{Gao2010,
author = {Gao, Daqiang and Yang, Guijin and Li, Jinyun and Zhang, Jing and Zhang, Jinlin and Xue, Desheng},
title = {Room-Temperature Ferromagnetism of Flowerlike {CuO} Nanostructures},
journal = {The Journal of Physical Chemistry C},
volume = {114},
number = {43},
pages = {18347-18351},
year = {2010},
doi = {10.1021/jp106015t},
URL = {https://doi.org/10.1021/jp106015t},
}

@article{Soon2009,
  title = {Native defect-induced multifarious magnetism in nonstoichiometric cuprous oxide: First-principles study of bulk and surface properties of {Cu}$_{2\ensuremath{-}\ensuremath{\delta}}\text{O}$},
  author = {Soon, Aloysius and Cui, Xiang-Yuan and Delley, Bernard and Wei, Su-Huai and Stampfl, Catherine},
  journal = {Phys. Rev. B},
  volume = {79},
  issue = {3},
  pages = {035205},
  numpages = {15},
  year = {2009},
  month = {01},
  publisher = {American Physical Society},
  doi = {10.1103/PhysRevB.79.035205},
  url = {https://link.aps.org/doi/10.1103/PhysRevB.79.035205}
}

@article{Bogenrieder2024,
author = {Bogenrieder, Stefanie E. and Beßner, Julian and Engstfeld, Albert K. and Jacob, Timo},
title = {First-Principles Study on the Structural and Magnetic Properties of Low-Index {Cu}$_2${O} and {CuO} Surfaces},
journal = {The Journal of Physical Chemistry C},
volume = {128},
number = {23},
pages = {9693-9704},
year = {2024},
doi = {10.1021/acs.jpcc.4c01102},
URL = {https://doi.org/10.1021/acs.jpcc.4c01102},
}

@article{Kumar2012,
  title = {Detection of Vibration-Mode Scattering in Electronic Shot Noise},
  author = {Kumar, Manohar and Avriller, R\'emi and Yeyati, Alfredo Levy and van Ruitenbeek, Jan M.},
  journal = {Phys. Rev. Lett.},
  volume = {108},
  issue = {14},
  pages = {146602},
  numpages = {4},
  year = {2012},
  month = {04},
  publisher = {American Physical Society},
  doi = {10.1103/PhysRevLett.108.146602},
  url = {https://link.aps.org/doi/10.1103/PhysRevLett.108.146602}
}

@article{Blanter2000,
title = {Shot noise in mesoscopic conductors},
journal = {Physics Reports},
volume = {336},
number = {1},
pages = {1-166},
year = {2000},
issn = {0370-1573},
doi = {https://doi.org/10.1016/S0370-1573(99)00123-4},
url = {https://www.sciencedirect.com/science/article/pii/S0370157399001234},
author = {Ya.M. Blanter and M. Büttiker}
}

@article{Batsaikhan2020,
author = {Batsaikhan, Erdembayalag and Lee, Chi-Hung and Hsu, Han and Wu, Chun-Ming and Peng, Jen-Chih and Ma, Ma-Hsuan and Deleg, Sangaa and Li, Wen-Hsien},
title = {Largely Enhanced Ferromagnetism in Bare {CuO} Nanoparticles by a Small Size Effect},
journal = {ACS Omega},
volume = {5},
number = {8},
pages = {3849-3856},
year = {2020},
doi = {10.1021/acsomega.9b02913},
URL = {https://doi.org/10.1021/acsomega.9b02913},
}

@article{Prestel2019,
  title = {Tuning the magnetic anisotropy energy of atomic wires},
  author = {Prestel, Martin W. and Ritter, Markus F. and Di Bernardo, Angelo and Pietsch, Torsten and Scheer, Elke},
  journal = {Phys. Rev. B},
  volume = {100},
  issue = {21},
  pages = {214439},
  numpages = {6},
  year = {2019},
  month = {12},
  publisher = {American Physical Society},
  doi = {10.1103/PhysRevB.100.214439},
  url = {https://link.aps.org/doi/10.1103/PhysRevB.100.214439}
}

@article{Calvo2012,
  title = {Analysis of the {Kondo} effect in ferromagnetic atomic-sized contacts},
  author = {Calvo, M. R. and Jacob, D. and Untiedt, C.},
  journal = {Phys. Rev. B},
  volume = {86},
  issue = {7},
  pages = {075447},
  numpages = {17},
  year = {2012},
  month = {08},
  publisher = {American Physical Society},
  doi = {10.1103/PhysRevB.86.075447},
  url = {https://link.aps.org/doi/10.1103/PhysRevB.86.075447}
}

@article{Calvo2009,
  title={The {Kondo} effect in ferromagnetic atomic contacts},
  author={Calvo, M Reyes and Fernandez-Rossier, Joaquin and Palacios, Juan Jos{\'e} and Jacob, David and Natelson, Douglas and Untiedt, Carlos},
  journal={Nature},
  volume={458},
  number={7242},
  pages={1150--1153},
  year={2009},
  publisher={Nature Publishing Group UK London},
  doi = {10.1038/nature07878},
  url = {https://doi.org/10.1038/nature07878}
}

@article{Strigl2016,
  title = {Magnetism in Pd: Magnetoconductance and transport spectroscopy of atomic contacts},
  author = {Strigl, F. and Keller, M. and Weber, D. and Pietsch, T. and Scheer, E.},
  journal = {Phys. Rev. B},
  volume = {94},
  issue = {14},
  pages = {144431},
  numpages = {12},
  year = {2016},
  month = {10},
  publisher = {American Physical Society},
  doi = {10.1103/PhysRevB.94.144431},
  url = {https://link.aps.org/doi/10.1103/PhysRevB.94.144431}
}

@article{Strohmeier2023,
    author = {Strohmeier, Marcel and Kirchberger, Kim and Scheer, Elke},
    title = {Nonlinear transport properties of atomic copper point contacts},
    journal = {Low Temperature Physics},
    volume = {49},
    number = {7},
    pages = {827-833},
    year = {2023},
    month = {07},
    issn = {1063-777X},
    doi = {10.1063/10.0019693},
    url = {https://doi.org/10.1063/10.0019693},
}


\newpage

\section*{Figures and Figure Captions}

\large{TOC graphic}

\begin{figure}[ht!]
    \centering
   \includegraphics[width=8.2cm]{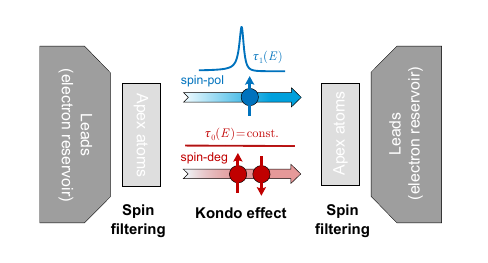}
   \end{figure}
\newpage

\large{Figure 1}

\begin{figure}[ht!]
    \centering
    \includegraphics[width=1\columnwidth]{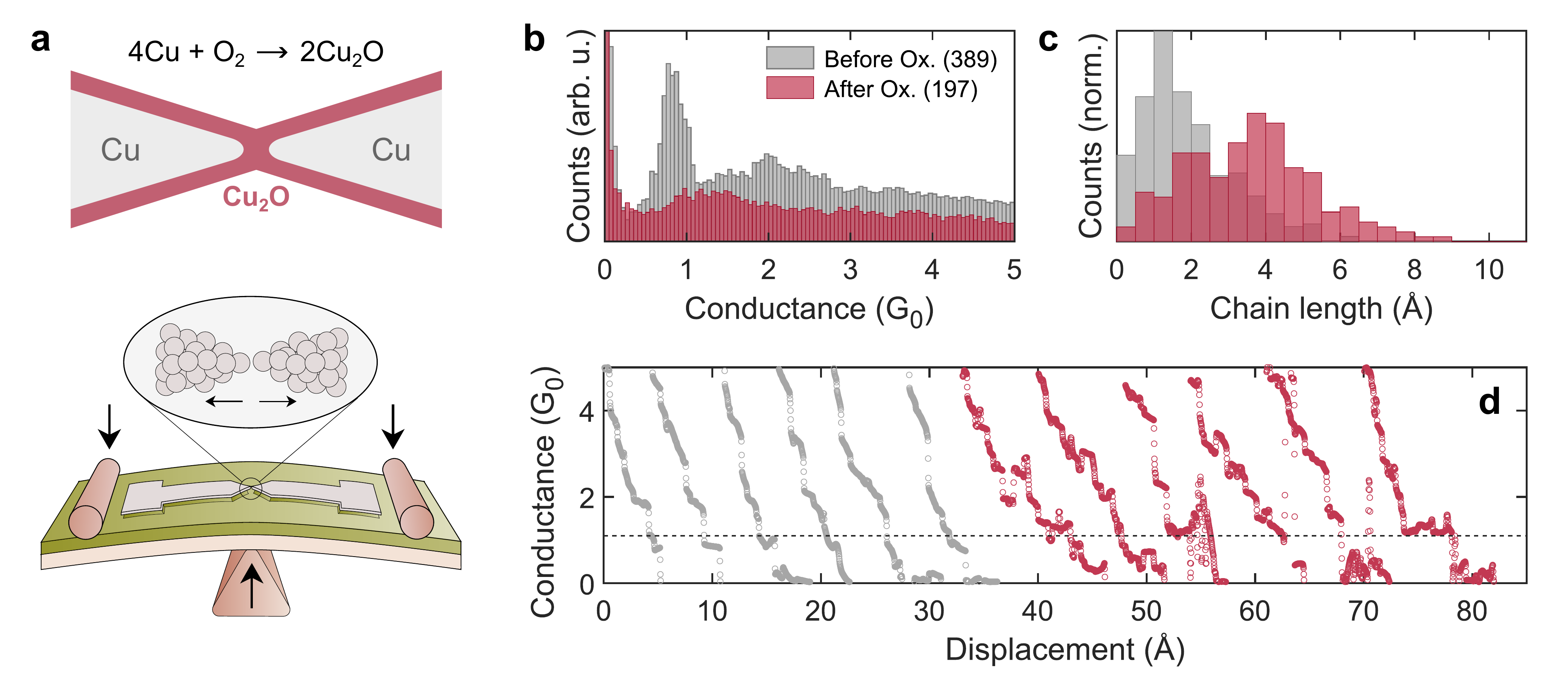}
    \caption{Analysis of the opening traces recorded on an air oxidized Cu MCBJ, comparing measurements taken during two consecutive cool-downs. (a) Cartoon of the oxidized atomic contact. The gray areas indicate the metallic Cu, the reddish the oxidized areas at the surface of the leads and in the contact region. (b) Conductance histogram constructed from all opening curves. (b) Distribution of chain lengths extracted from the traces. (c) Examples of individual opening traces. Gray/red: First cool-down (without/with breaking the junction at room temperature)}
    \label{fig:Histogram}
\end{figure}

\newpage
\large{Figure 2}

\begin{figure*}[ht!]
    \centering
    \includegraphics[width=1\columnwidth]{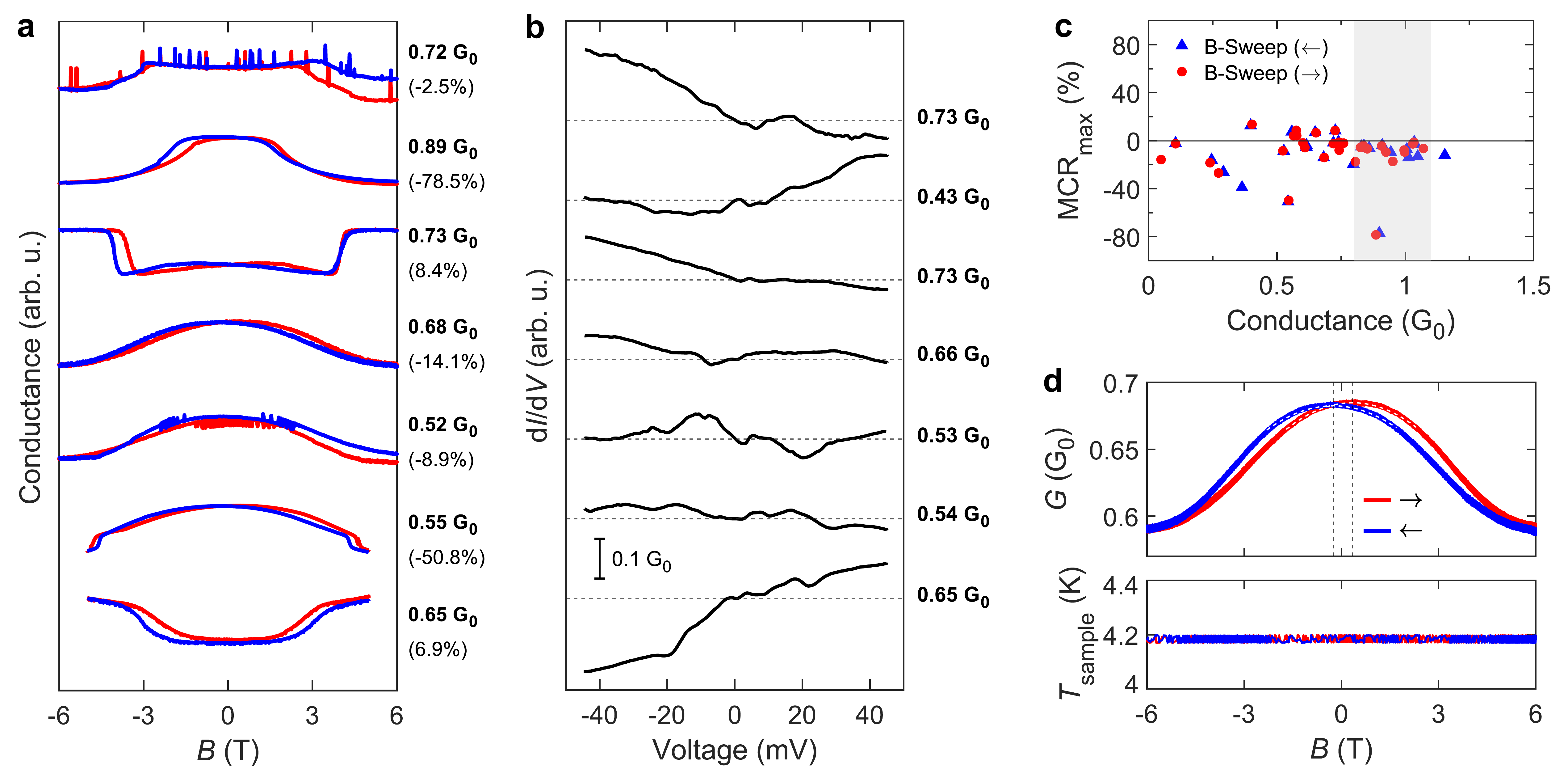}
    \caption{Magnetoconductance (MC) measurements on atomic Cu/Cu$_x$O contacts. (a) Representative non-monotonic MC traces recorded after ambient-air exposure in a semi-broken state. Curves are scaled and vertically off-set for clarity.  
    In all measurements, the magnetic field is applied perpendicular to the sample plane and swept between $\pm5$\,T or $\pm6$\,T. Blue (red) traces denote the return (forward) sweep. Labels indicate the zero-bias conductance (bold) and the maximum magnetoconductance ratio MCR$_\text{max}$ (in parentheses). (b) Corresponding $I$–$V$ characteristics measured prior to each MC sweep and vertically off-set (see conductance scale bar). (c) MCR$_\text{max}$ as a function of zero-field conductance for all continuous MC sweeps of the same sample; different symbols distinguish sweep directions. The MCR is mostly negative, but it can also be positive.
    (d) Example MC trace measured during a second cool-down after air exposure, showing nonlinear and hysteretic behavior at $G<1\,G_0$ (upper panel). The simultaneously recorded sample temperature remains constant throughout the field sweep (lower panel).     }
    \label{fig:MC_records}
\end{figure*}

\newpage
\large{Figure 3}

\begin{figure}[ht!]
    \centering
    \includegraphics[width=1\columnwidth]{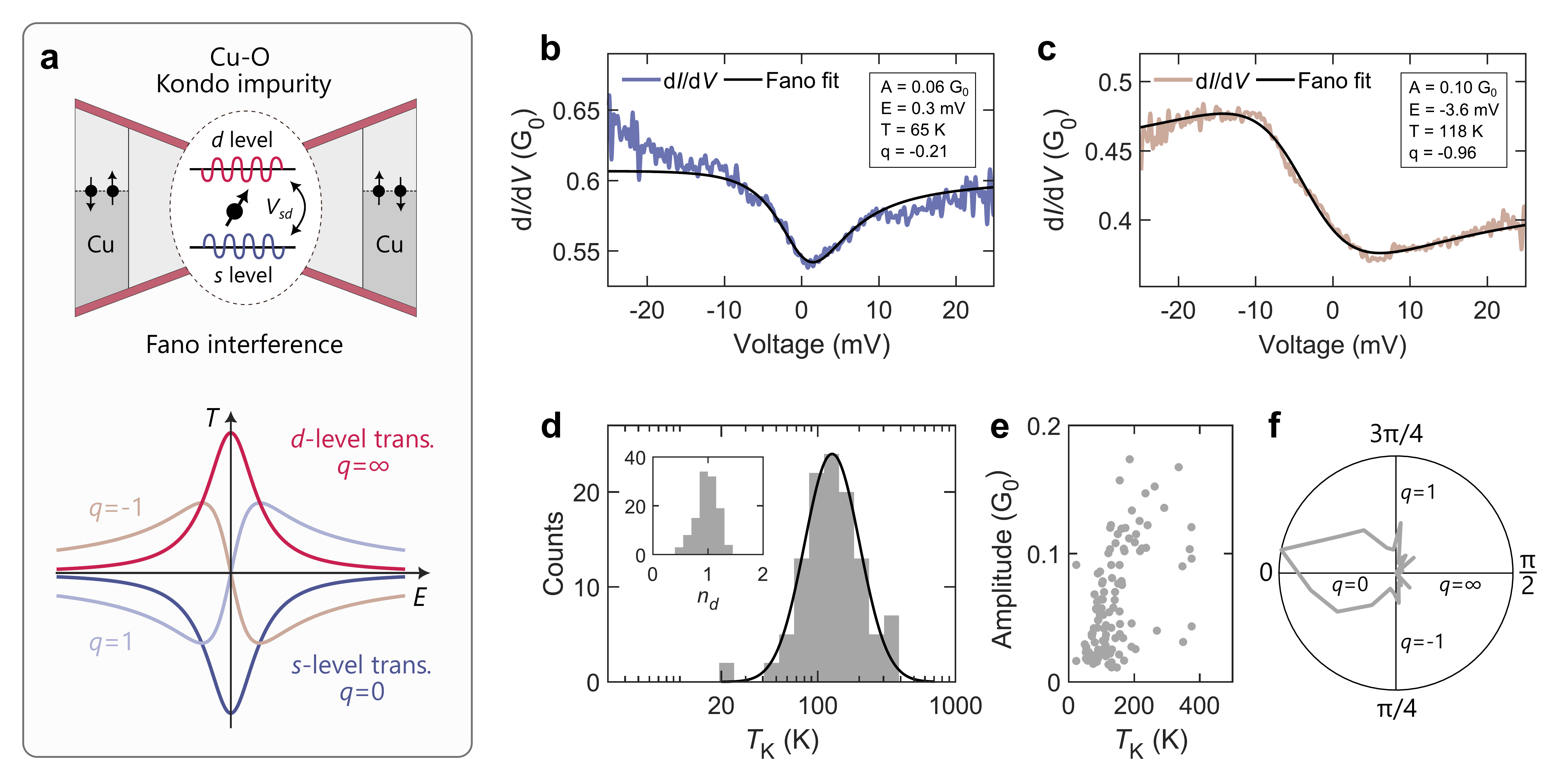}
    \caption{Analysis of the Kondo effect in an oxidized Cu MCBJ: (a) Sketch of an atomic contact featuring a single Cu-O Kondo impurity localized within its constriction. The Fano interference between the $s$- and $d$-levels, hybridized by $V_{sd}$, gives rise to an energy-dependent transmission $T(E)$ whose Fano shape depends on the $q$ factor. (b), (c) Examples of experimentally measured d$I$/d$V$ curves exhibiting Fano line shapes. A fit to Eq.~\eqref{eq:fano_line_shape} is shown by the solid line. (d) Histogram of the Kondo temperature $T_K$ (on a log scale) and occupation number $n_d$ (inset) across all contacts with $A>0.01\,G_0$. A log-normal distribution is fitted to the $T_K$ data. (e) Scatter plot of $A$ versus $T_K$. (f) Polar plot with $\alpha = \arctan(q)$, representing the distribution of the shape factor $|q| \in [0,\,\infty)$.}
    \label{fig:Kondo_MS_EM2_09}
\end{figure}

\newpage
\large{Figure 4}
\begin{figure}[ht!]
    \centering
    \includegraphics[width=1.0\columnwidth]{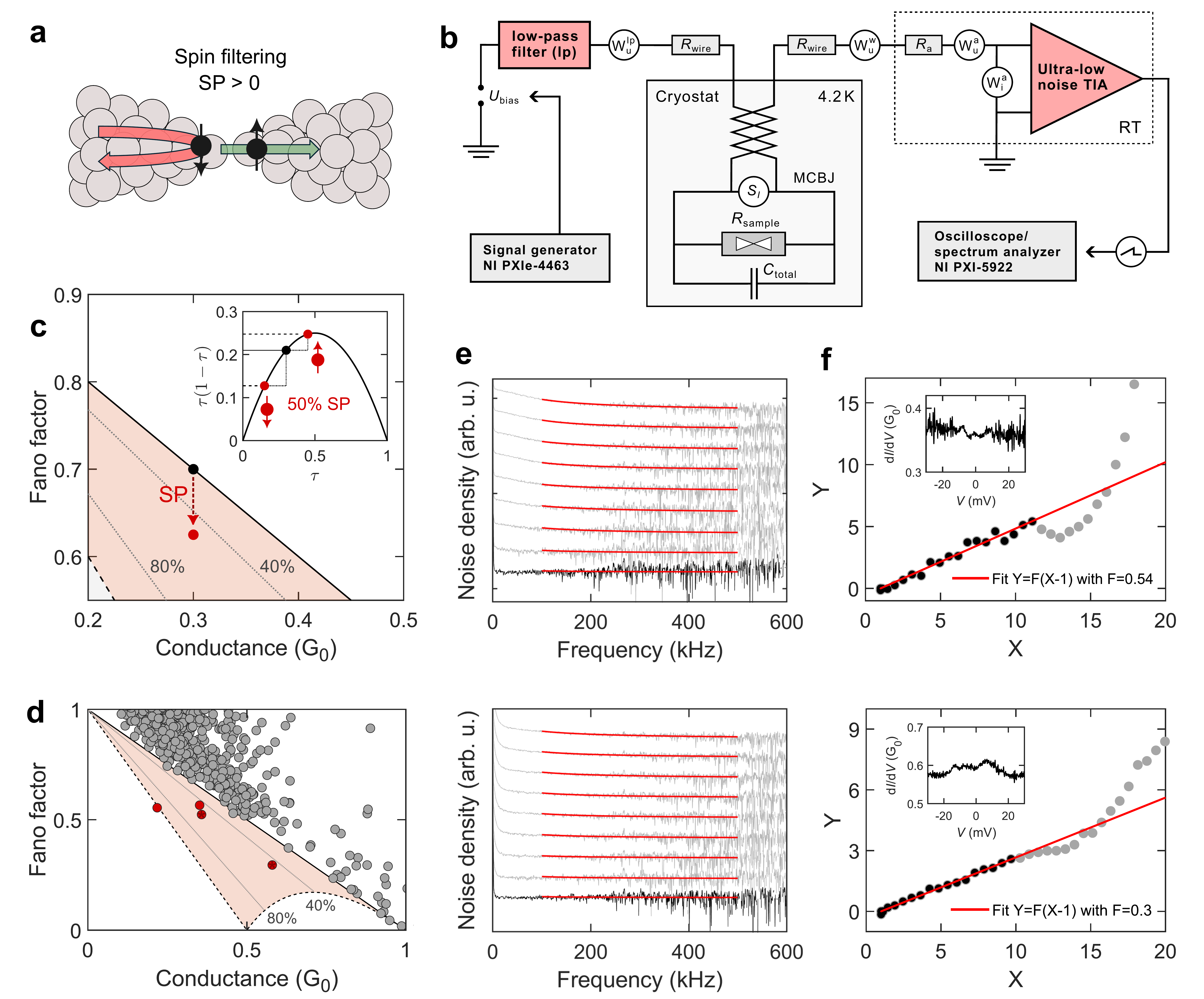}
    \caption{(a) A Cartoon depicting the spin selectivity of an atomic Cu$_x$O junction. (b) Circuit diagram of the electronic setup for measuring shot noise at low temperatures. (c) Illustration of the Fano factor reduction induced by the spin polarization (here: SP$=50\%$) of a single conductance channel. Inset: Functional evolution of the partition noise expression $\tau (1-\tau)$ versus $\tau$; the black dot denotes the spin-degenerate case with $\tau=0.3$, while red dots represent the spin-polarized channel configuration with equal total transmission, corresponding to the marked Fano factors in the main plot.  (d) Fano factor analysis in the $F(G)$ space, with the red data points highlighting contacts entering the spin polarized transport regime (red area). The noise spectra for two contact with SP along with fits to Eq.~\eqref{eq:1_f_fit} are shown in (e). Black spectra correspond to the zero-bias (thermal) noise, while gray spectra are obtained under non-equilibrium conditions with increasing bias from bottom to top and shifted vertically for clarity. 
    The plots (f) display the noise-voltage characteristics in reduced units $X$ and $Y$, together with a linear fit to the function $Y = F(X - 1)$ to extract the Fano factor $F$ of each contact, respectively. The d$I$/d$V$ curves are shown in the insets. }
    \label{fig:noise_spinpol_contacts}
\end{figure}

\newpage
\large{Figure 5}
\begin{figure*}[ht!]
    \centering
    \includegraphics[width=1.0\columnwidth]{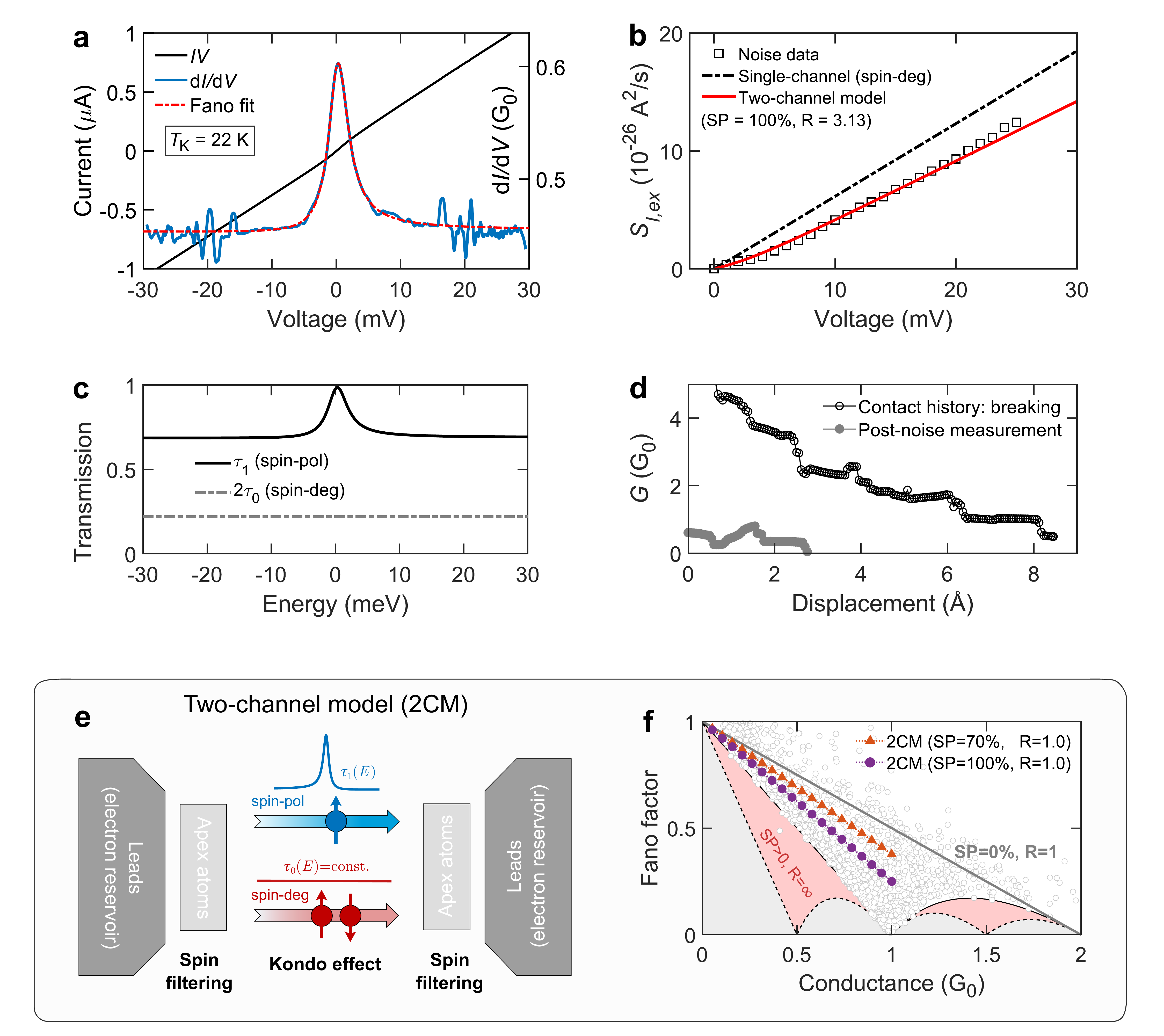}
    \caption{Noise modeling on an atomic contact exhibiting a strong Kondo resonance: (a) $IV$ and d$I$/d$V$ characteristics, together with the Fano fit used to extract $T(E)$. (b) Excess noise $S_{I,\text{ex}}(V)$ compared with noise modeling that includes an energy dependent transmission. Calculations for the spin degenerate single-channel model (dash-dotted line) and the 2CM (red line) are shown. (c) Transmission functions underlying the 2CM in (b) with $SP=100\%$ and $R=3.13$ (d) Conductance versus displacement before and after the noise measurement. (e) Sketch of the 2CM applied to a Cu$_x$O atomic junction: A spin degenerate channel with a bias independent transmission $\tau_0$, together with a spin polarized channel, defines the transmission of the monoatomic constriction. The polarized channel exhibits an energy dependent transmission function $\tau_1(E)$ and carries the profile of the Kondo resonance, which is induced by the magnetic moments of the Cu-O units. Additional spin filtering may occur in the nanoscale vicinity of the apex atoms, where surface magnetism in oxidized Cu becomes more dominant. (f) $F(G)$ plot interpreted within the 2CM. Two exemplary lines with fixed values of $R$ and $SP$ are indicated by colored marker symbols.}
    \label{fig:noise_model}
\end{figure*}

\newpage
\large{Figure 6}
\begin{figure}[ht!]
    \centering
    \includegraphics[width=0.5\columnwidth]{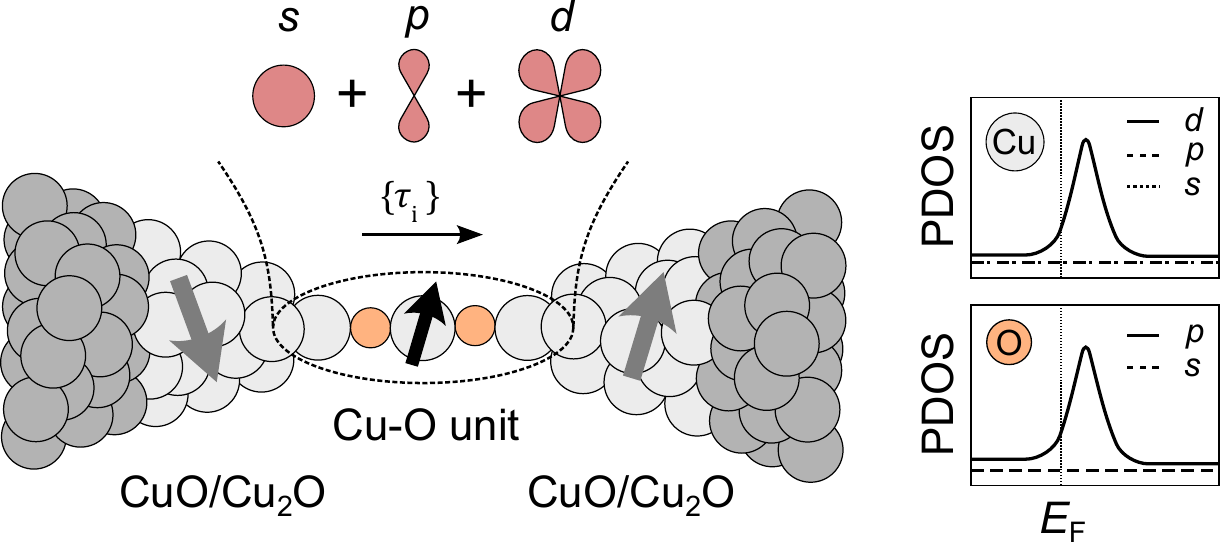}
    \caption{Sketch of a Cu$_x$O atomic junction: The transmission properties are governed by a monoatomic chain of Cu-O units bridging the two electrodes. The interplay of all contributing orbitals gives rise to a net magnetic moment and spin polarized currents, predominantly hosted by the $d$ and $p$ orbitals 
    of the Cu and O atoms, respectively. Surface oxidation at the outermost apex atoms may induce a finite magnetization that affects the transport properties. The projected density of states (PDOS) is cartooned following \autocite{Zheng2015}.}
    \label{fig:drawing_contact_model}
\end{figure}

\end{document}


\maketitle








\tableofcontents

\section{Experimental}

\noindent The experiments were carried out using mechanically controlled thin-film break junctions (thin-film MCBJs). We fabricate the samples by electron beam lithography and electron beam evaporation in lift-off technique in combination with a subsequent oxygen plasma treatment to define narrow suspended Cu bridges on an insulating substrate. The purity of the starting material is Cu $>$99.999\,\% with a Fe content of 0.01\,ppm. The thickness and width of the pristine nanowire are about 100\,nm, guaranteeing a metallic core despite the plasma treatment \cite{Ruitenbeek1996, Strigl2015}. 
\\
All measurements were carried out in cryogenic vacuum at a base temperature of 4.2\,K. To reach these environmental conditions, two complementary measurement inserts were used in a helium bath (wet) cryostat: one dedicated to magnetotransport measurements and the other one specifically designed for shot noise experiments \cite{Strigl2015, Prestel2021}. The noise measurement setup and how to extract the Fano factor is described in the SI.
\\
Oxidation was achieved via passive exposure to ambient air, combined with a controlled pre-cooling procedure designed to enhance oxygen incorporation into the nano-constriction. Specifically, pristine Cu junctions were first ruptured in air into the tunneling regime and subsequently re-closed before cooling down. This protocol promotes surface oxidation at the rupture site and increases the likelihood that subsequent low-temperature breaks occur within the oxidized region.

\section{Calibration of the electrode displacement}

\begin{figure}[h!]
    \centering
    \includegraphics[width=0.8\columnwidth]{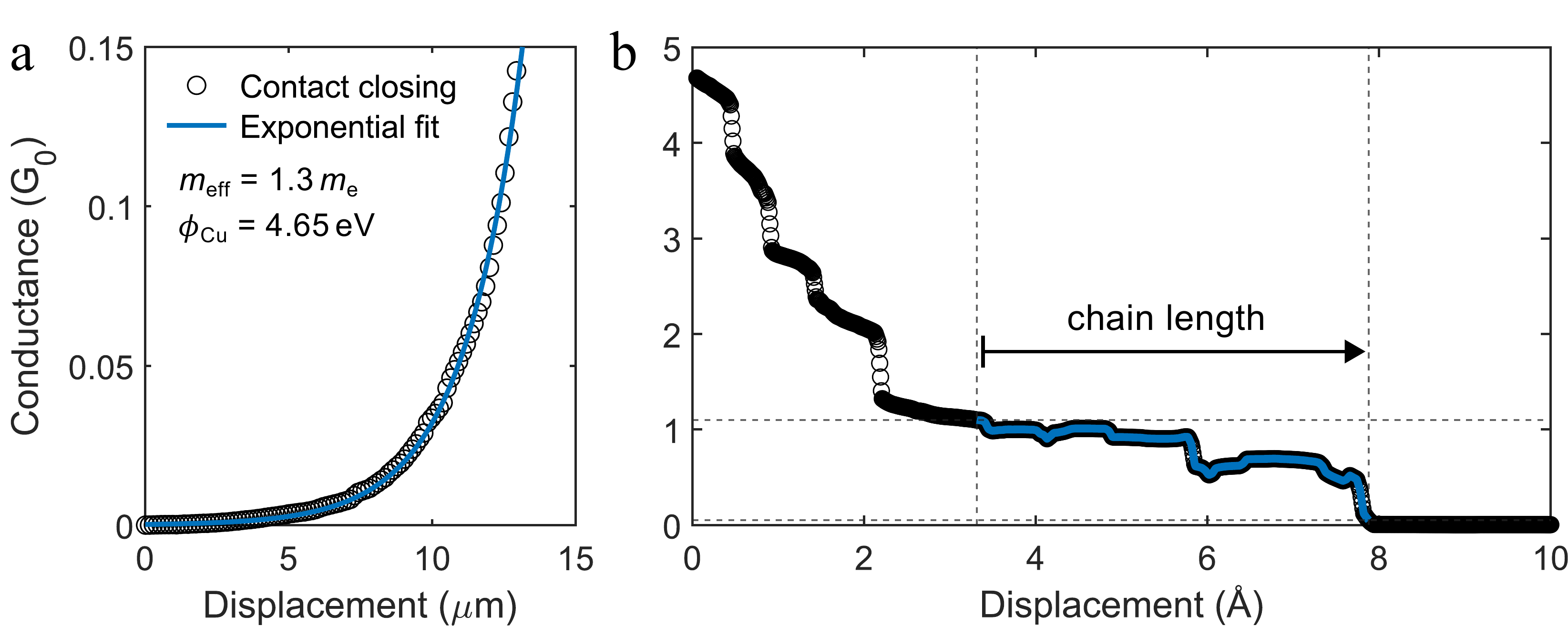}
    \caption{(a) Exponential increase of the tunnel conductance when closing the break junction, leading to the formation of an atomic-size contact. Fitting the trace with respect to the displacement (solid line) can be used to calculate the reduction ration $r$, necessary for the calibration of the horizontal displacement of the junction. (b) Example of an opening trace illustrating the definition of the chain length used in the analysis: as indicated by the dashed lines, the length is defined as the displacement between the two points where the conductance drops below 1.1 and 0.05\,$G_0$, respectively. After calibration, the conductance trace attributed to the atomic-size chain (bluish solid line) typically evolves over elongations on the order of a few angstroms.}
    \label{fig:chain_length_def}
\end{figure}

\noindent The determination of atomic chain lengths requires a calibration of the mechanical displacement in the break junction experiment. The bending mechanics produces a vertical substrate displacement $\Delta z$, which translates into a horizontal elongation of the junction $\Delta u$. The relation between both quantities is described by the reduction ratio $r=\Delta u / \Delta z$ \cite{Ruitenbeek1996}.

The calibration of $r$ uses the exponential dependence of the tunneling conductance on the electrode separation. When the junction approaches the closure, the conductance follows an exponential functional evolution
\begin{equation}
    G \sim \exp{\left( \dfrac{2}{\hbar}\sqrt{2 m_\text{eff}\phi}\cdot \Delta u \right)},
\end{equation}
where $m_\text{eff}$ denotes the effective electron mass and $\phi$ the work function of the material. Expressing the elongation through $\Delta u=r\,\Delta z$ yields the fit function
\begin{equation}
    G(\Delta z) = A \exp{\left( \dfrac{2}{\hbar}\sqrt{2 m_\text{eff}\phi}\, r\, \Delta z \right)},
\end{equation}
with $\Delta z$ given in units of $\mu$m. A representative fit to an experimental trace is shown in Fig.~\ref{fig:chain_length_def} (a). Due to the absence of reliable parameters for the Cu$_x$O junction, both $m_\text{eff}$ and $\phi$ are approximated by the values reported for Cu (see inset). Closing traces may show deviations from a purely exponential behavior close to $1\,G_0$, which likely originate from structural rearrangements in Cu$_x$O configurations. These deviations are not included in the fit. Instead, the reduction ratio $r$ is exclusively verified on pristine traces recorded on a particular sample.\\

After calibration, the horizontal electrode displacement can be expressed in absolute units. Fig. \ref{fig:chain_length_def} (b) shows an example of a calibrated opening trace. The elongation of an atomic contact typically occurs over a displacement of a few angstroms. The chain length is defined following the procedure introduced by Thijssen \textit{et al.}~\cite{Thijssen2008}. For each opening trace, the two positions at which the conductance drops below $1.1\,G_0$ and $0.05\,G_0$, respectively, are determined. The displacement between these points defines the chain length. Within this interval (indicated by the black arrow) the conductance often exhibits a non-monotonic evolution that consists of several discrete jumps. These features reflect structural rearrangements during the final stage of the chain elongation.


\section{Noise measurement setup and modeling}
\label{chap:noise_setup}

\begin{figure}[t!]
    \centering
    \includegraphics[width=0.75\columnwidth]{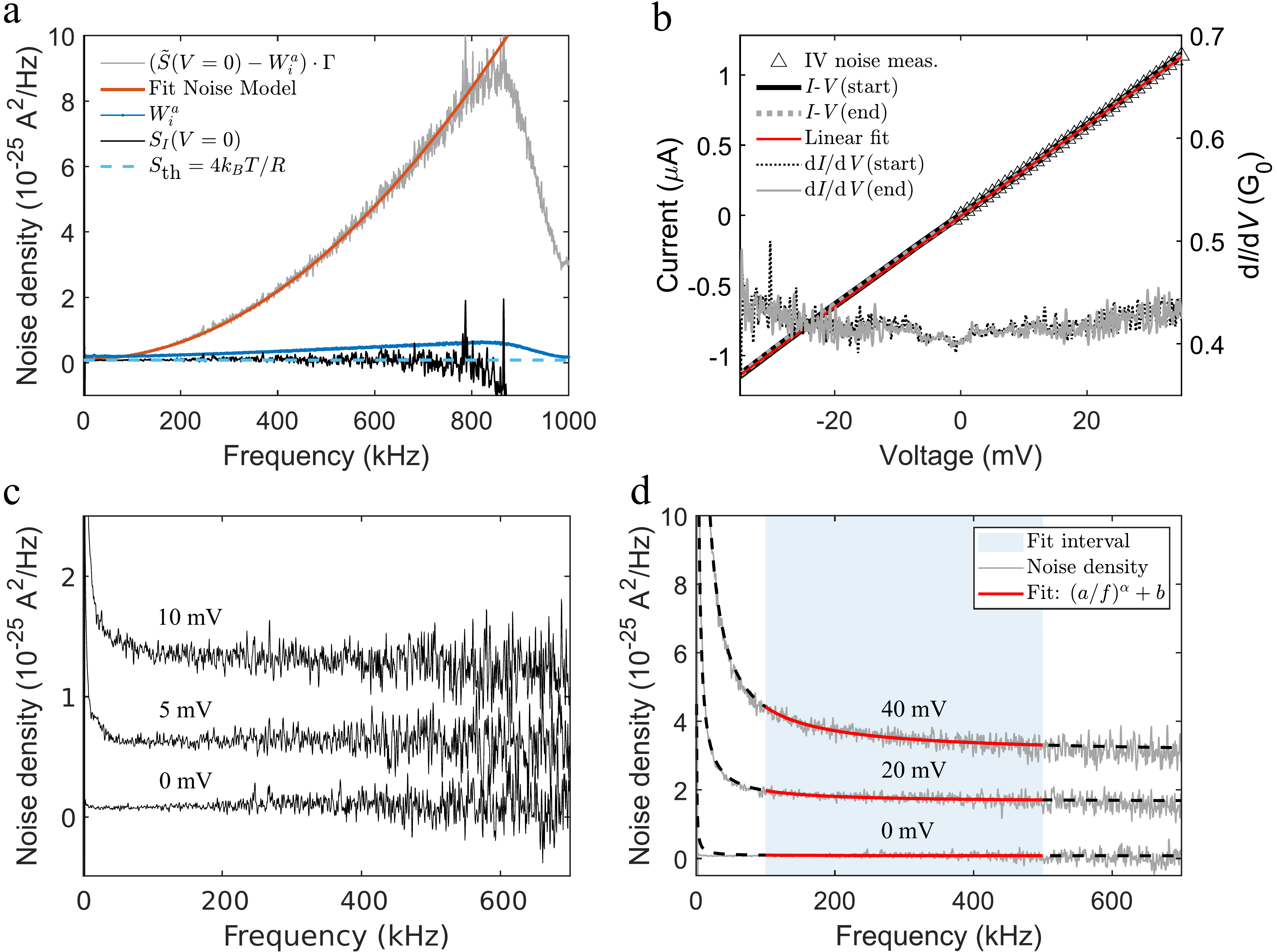}
    \caption{Noise modeling and data postprocessing for one particular atomic contact: (a) Measured  noise $(\Tilde{S}(V=0)-W_\text{i}^\text{a})\,\Gamma$ (gray) and fit to the noise model of Eq. \eqref{eq:noise_model} (red) using $T$ and $C_\text{total}$ as free parameters. A correction according to Eq. \eqref{eq:noise_model} yields $S_I(V=0)$ (black), the sample's intrinsic current noise. At zero bias, $S_I$ is equals the thermal noise $S_\text{th}$ (light-blue dashed), which fall in the same range as $W_i^a$ (blue). (b) $I$-$V$ characteristics and numerically computed d$I$/d$V$s measured prior and subsequent to the noise measurement to determine $R_\text{sample}$. (c) Corrected spectra $S_I(V)$ at varying bias voltage. (d) Fitting of the spectra $S_I(f)$ with the function $(a/f)^\alpha + b$ to account for $1/f^\alpha$-type noise contributions, while extracting the white noise part $S_w := b$. The area shaded in blue indicates the fit interval.}
    \label{fig:noise_modeling}
\end{figure}

\noindent Accurate shot noise measurements on atomic-scale contacts require a detailed modeling of the measurement circuit. The recorded signal contains contributions from several instrumental noise sources in addition to the intrinsic current fluctuations of the sample. A reliable extraction of the sample noise therefore requires a quantitative description of the full measurement setup, including its impedance and frequency response. In the following, the dominant noise sources of the setup are summarized and incorporated into an equivalent circuit model.\\

Four principal noise sources contribute to the measured signal.\\

\noindent \textbf{(i) Low-pass filter (adder):}  
The home-built adder stage operates as a low-pass filter and is placed behind the voltage source (NI PXIe-4463). Both the adder itself and the voltage source contribute voltage noise contribute an effective voltage noise contribution.\\

\noindent \textbf{(ii) Transimpedance amplifier (TIA):}  
The TIA converts current fluctuations into a measurable voltage signal and introduces both current and voltage noise. The frequency dependent current noise $W_\text{i}^{\text{a}}$ can be obtained directly from a noise spectrum recorded with an open input.\\

\noindent \textbf{(iii) Wiring:}  
The entire cabling between the room temperature electronics and the cryogenic insert produces a thermal voltage noise.\\

\noindent \textbf{(iv) Sample (MCBJ):}  
The mechanically controllable break junction generates the intrinsic current noise $S_I$ of interest, which consists of thermal noise and bias dependent excess noise (shot noise).\\

Each noise contribution $W$ is represented as an independent voltage or current source in an equivalent circuit shown in Fig.~4 (a). Capacitances originating from the sample, wiring, and electronic components are combined into an effective capacitance on the order of $C_\text{total} \sim 100\,\text{pF}$. The sample resistance $R_\text{sample}$ acts in parallel with this capacitance, forming a frequency dependent impedance that can be written as
\begin{equation}
\frac{1}{Z'} = i \omega C_\text{total} + \frac{1}{R_\text{sample}} .
\end{equation}

This impedance appears in series with the remaining setup resistance $R_\text{setup} = R_\text{wire} + R_\text{amp}$, where $R_\text{wire}$ accounts for the resistance of the cabling and $R_\text{amp}$ represents the input resistance of the amplifier stage. The total impedance of the detection circuit therefore becomes
\begin{equation}
Z_\text{total} = Z' + R_\text{setup}.
\end{equation}

With this, voltage noise contributions originating from the electronic components (and listed above) are converted into a measurable current noise by the impedance of the circuit. Their corresponding contribution to the measured noise spectrum is given by
\begin{equation}
\Tilde{W}_\text{i}^{\text{setup}} =
\frac{W_\text{u}^{\text{lp}} + W_\text{u}^{\text{a}} + W_\text{u}^{\text{w}}}{|Z_\text{total}|^2}.
\end{equation}

At the same time, the intrinsic current noise $S_I$ generated by the sample is filtered by the same circuit elements. As a result, the measured sample noise $\Tilde{S}_I$ is related to $S_I$ via the relation
\begin{equation}
\Tilde{S}_I = S_I \left| \frac{Z'}{Z_\text{total}} \right|^2 ,
\end{equation}

\noindent which defines the transfer function of the measurement circuit
\begin{equation}
\mathcal{H} = \frac{\Tilde{S}_I}{S_I}=\left|\frac{Z'}{Z_\text{total}}\right|^2 .
\end{equation}

Taking all instrumental noise sources of the whole circuit into account yields the complete model based analytical expression for the measured noise spectrum
\begin{equation}
\Tilde{S} =
S_I\left|\frac{Z'}{Z_\text{total}}\right|^2
+
\frac{W_\text{u}^{\text{lp}}+W_\text{u}^{\text{a}}+W_\text{u}^{\text{w}}}{|Z_\text{total}|^2}
+
W_\text{i}^{\text{a}} .
\label{eq:noise_model}
\end{equation}

To extract the intrinsic sample noise, the spectrum measured at zero bias $\Tilde{S}(V=0)$ is used to initialize the model. At zero bias, the sample contribution reduces to the thermal noise $S_\text{th}=4k_BT/R_\text{sample}$. Eq. (\ref{eq:noise_model}) can therefore be fitted using the temperature $T$ and the total capacitance $C_\text{total}$ as free parameters. A fit to a representative spectrum $\Tilde{S}(V=0)$, along with the corrected spectra $S_I(V=0)$, $W_\text{i}$ and the corresponding thermal noise $S_\text{th}$ is illustrated in Fig. \ref{fig:noise_modeling} (a). To obtain $R_\text{sample}$, $IV$ curves are measured before and after each noise measurement [see Fig. \ref{fig:noise_modeling} (b)]. With the calibrated capacitance $C_\text{total}$, spectra recorded at finite bias can be corrected according to the relation
\begin{align}
S_I(f,V) = 
\Biggl[ (\Tilde{S}(f,V) - W_\text{i}^{\text{a}}(f))\,\Gamma(f) 
 - \frac{W_\text{u}^{\text{lp}} + W_\text{u}^{\text{a}}(f) + W_\text{u}^{\text{w}}}{|Z_\text{total}|^2} \Biggr] 
\left| \frac{Z'}{Z_\text{total}} \right|^{-2} ,
\end{align}

\noindent as depicted in Fig. \ref{fig:noise_modeling} (c). Note that within the Landauer-Büttiker framework thermal noise and shot noise should exhibit a white spectral density. Atomic contacts, however, may display additional frequency dependent contributions. Enhanced $1/f^\alpha$ noise in atomic junctions has been reported, particularly at elevated bias voltages \cite{Mu2019}. To separate white noise from the frequency dependent component, each corrected spectrum is fitted with
\begin{equation}\label{eq:1/f_fit}
    S_I(f) = \left(\frac{a}{f}\right)^\alpha + b ,
\end{equation}
where $a$, $b$, and $\alpha$ are free parameters [see Fig. \ref{fig:noise_modeling} (d)]. The constant term $b$ defines the white-noise contribution $S_w := b$, which is used for the subsequent shot noise analysis. The fitting interval typically spans $f \in [100,500]\,\text{kHz}$, where the measurement bandwidth remains free from instrumental cutoff effects.

\section{Additional examples of Kondo active junctions}

\begin{figure}[t!]
    \centering
    \includegraphics[width=0.75\columnwidth]{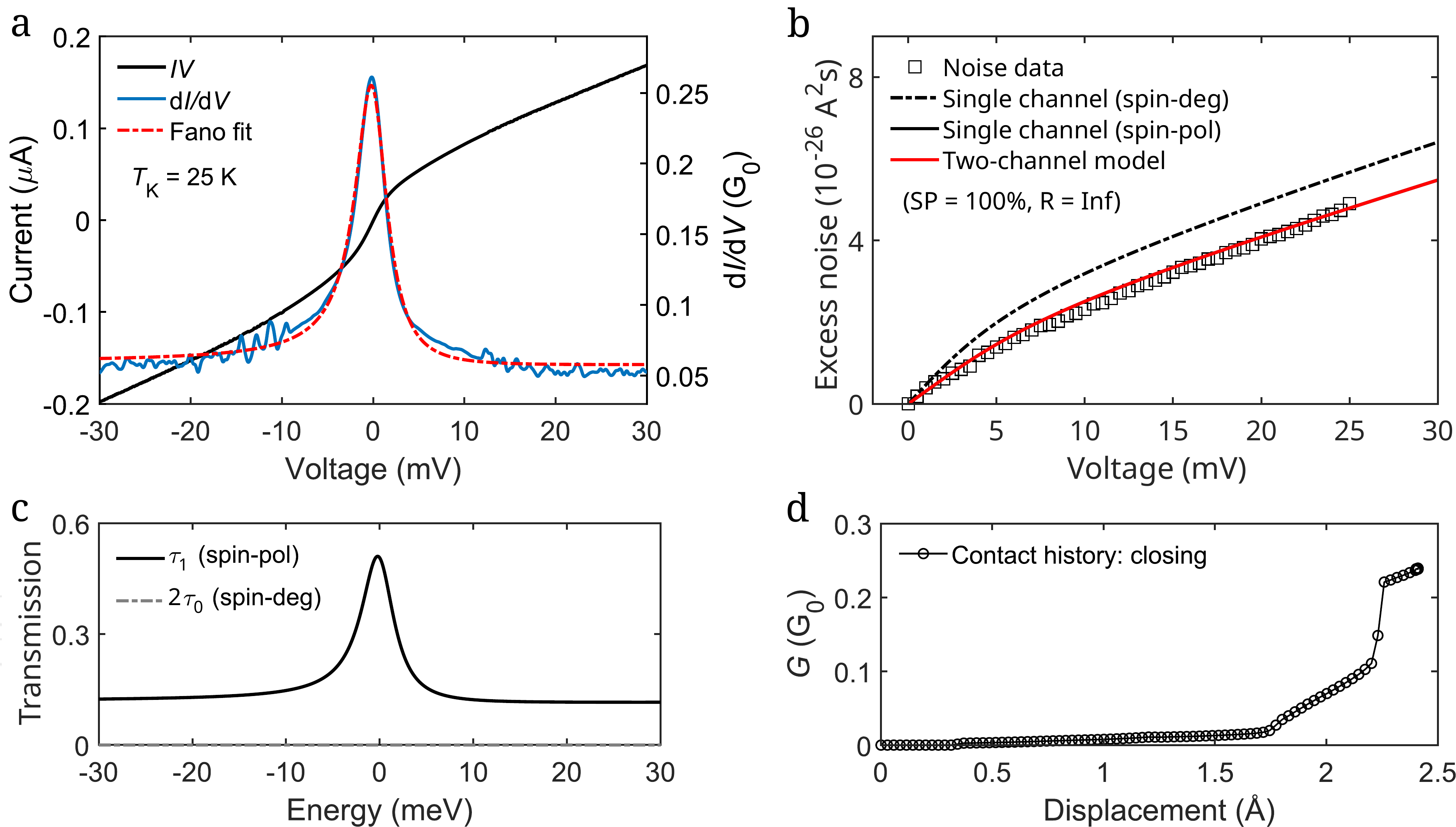}
    \caption{Noise modeling on contact 2 exhibiting a strong Kondo resonance: (a) $IV$ and d$I$/d$V$ characteristics, together with the Fano fit used to extract $\tau(E)$. (b) Excess noise $S_{I,\text{ex}}(V)$ compared with noise modeling that includes an energy dependent transmission. Calculations for the spin degenerate single-channel model (dash-dotted line) and the 2CM (red line) are shown. (c) Transmission functions underlying the 2CM. (d) Conductance versus displacement before and after the noise measurement.}
    \label{fig:Kondo_14-U5-06}
    \vspace{1cm}

    \centering
    \includegraphics[width=0.75\columnwidth]{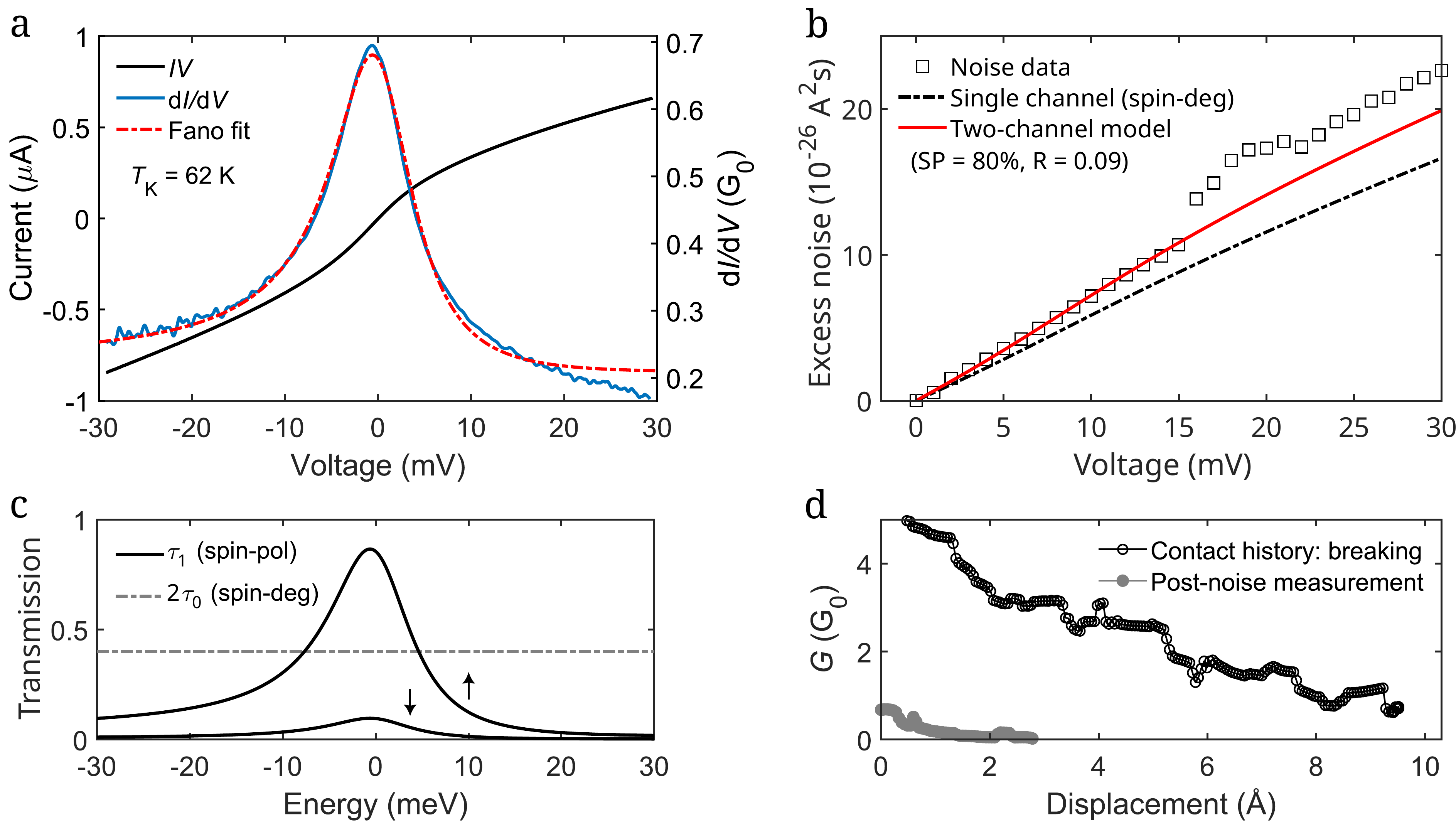}
    \caption{Noise modeling on contact 3 exhibiting a strong Kondo resonance: (a) $IV$ and d$I$/d$V$ characteristics, together with the Fano fit used to extract $\tau(E)$. (b) Excess noise $S_{I,\text{ex}}(V)$ compared with noise modeling that includes an energy dependent transmission. Calculations for the single-channel models (dash-dotted and black solid line) and the 2CM (red line) are shown. (c) Transmission functions underlying the 2CM. (d) Conductance versus displacement before and after the noise measurement.}
    \label{fig:Kondo_07-U7-68}
\end{figure}

\noindent In Figs. \ref{fig:Kondo_14-U5-06} and \ref{fig:Kondo_07-U7-68} we present further noise and transport measurements on atomic-scale Cu$_x$O junctions exhibiting pronounced Kondo resonances. In all cases, the (anomalous) excess noise is similarly compared with the noise modeling (2CM) as described in the main text. 

\section{Two-channel noise modeling}

\noindent The formal description of the anomalous shot noise follows the Landauer formalism and is partly motivated by the work of Tewari \emph{et al.} \cite{Tewari2018} and Mu \emph{et al.} \cite{Mu2019}. The starting point is the energy dependent transmission function $T(E)$ extracted from the differential conductance using the Fano fit described in the main text. Within the 2CM this transmission is decomposed into a spin polarized channel $\tau_1(E)$ and a spin degenerate background channel $\tau_0$. The spin resolved transmission can therefore be written as
\begin{align}
\label{eq:transmission_spin_resolved}
T(E)
&= \sum_{i,\sigma}\tau_{i,\sigma}(E) = \tau_{0,\uparrow}+\tau_{0,\downarrow}+\tau_{1,\uparrow}(E)+\tau_{1,\downarrow}(E) \\
&=2\tau_0 + \tau_1(E)
\end{align}

with $\tau_{0,\uparrow}=\tau_{0,\downarrow}=\tau_0$. The polarized channel is expressed through the spin polarization parameter SP,
\begin{align}
\tau_{1,\uparrow}(E) &= \left(\frac{1+SP}{2}\right)\tau_1(E), \\
\tau_{1,\downarrow}(E) &= \left(\frac{1-SP}{2}\right)\tau_1(E).
\end{align}

In the zero-temperature limit the current follows from the Landauer expression

\begin{equation}
I(V)=\frac{e}{h}
\int_{-eV/2}^{eV/2}T(E)\,\mathrm{d}E
=
\frac{e}{h}
\int_{-eV/2}^{eV/2}
\left[\tau_1(E)+2\tau_0\right]\mathrm{d}E .
\end{equation}

The corresponding shot noise is obtained by summing the partition noise contributions of all spin channels,

\begin{align}
S_{I,\text{2CM}}(V)
&=
G_0
\sum_{i,\sigma}
\int_{-eV/2}^{eV/2}
\tau_{i,\sigma}(E)\left[1-\tau_{i,\sigma}(E)\right]\,\mathrm{d}E .
\end{align}

Substituting the spin resolved transmissions yields

\begin{align} S_{I,\text{2CM}}(V) &= G_0 \int_{-eV/2}^{eV/2} \left(\frac{1+SP}{2}\right)\tau_1(E) \left[1-\left(\frac{1+SP}{2}\right)\tau_1(E)\right] \,\mathrm{d}E \nonumber\\ &+ G_0 \int_{-eV/2}^{eV/2} \left(\frac{1-SP}{2}\right)\tau_1(E) \left[1-\left(\frac{1-SP}{2}\right)\tau_1(E)\right] \,\mathrm{d}E \nonumber\\ &+ 2eVG_0\,\tau_0(1-\tau_0). \end{align}

The parameters $SP$ and $R$ are defined in Eqs.~(11) and (12) in the main text. For a given parameter pair, $\tau_0$ and $\tau_1(E)$ follow from the condition
\begin{equation}
    T(E)=2\tau_0+\tau_1(E).
\end{equation}

The calculated noise $S_{I,\text{2CM}}(V)$ is then compared with the experimentally measured excess noise. The transmission ratio $R$ determines the relative contribution of the two channels at large bias
\begin{equation}  
R=\frac{\tau_1(E\gg k_B T_K)}{2\tau_0}.
\end{equation}

Two limiting regimes can therefore be distinguished
\begin{align*}
    R &< 1\,:\hspace{0.3cm} \text{if}\;\tau_0\;\text{dominates }\;G(|V|\gg k_B T_K) \\ 
    R &> 1\,:\hspace{0.3cm} \text{if}\;\tau_1\;\text{dominates }\;G(|V|\gg k_B T_K) 
\end{align*}

These limits provide a convenient way to interpret the transport characteristics of the junction within the suggested two-channel description for atomic-scale Cu$_x$O junctions.

\printbibliography